\documentclass[fleqn,usenatbib]{mnras}

\usepackage{newtxtext,newtxmath}
\usepackage[T1]{fontenc}

\DeclareRobustCommand{\VAN}[3]{#2}
\let\VANthebibliography\thebibliography
\def\thebibliography{\DeclareRobustCommand{\VAN}[3]{##3}\VANthebibliography}

\usepackage{graphicx}	% Including figure files
\usepackage{amsmath}	% Advanced maths commands
\usepackage{float}      % To indicate figure position
\usepackage{dblfloatfix}% To enable figures at the bottom of page
\usepackage{esint}      % For closed curve integrals

\title[Secular evolution of MMRs in the PER3BP]{Secular evolution of resonant small bodies: semi-analytical approach for arbitrary eccentricities in the coplanar case}

\author[J. Pons and T. Gallardo]{
Juan Pons,$^{1}$\thanks{E-mail: juan.pons.93@gmail.com}
and Tabaré Gallardo$^{1}$
\\
$^{1}$Departamento de Astronomía, Instituto de Física, Facultad de Ciencias, Universidad de la República, Iguá 4225, Montevideo 11400, Uruguay.\\
}

\date{Accepted XXX. Received YYY; in original form ZZZ}

\pubyear{2021}

\begin{document}
\label{firstpage}
\pagerange{\pageref{firstpage}--\pageref{lastpage}}
\maketitle

% Abstract of the paper
\begin{abstract}

We study the secular evolution of a particle in deep mean motion resonance (MMR) with a planet in the planar elliptic restricted three body problem. We do not consider any restriction neither in the planet's eccentricity $e_p$ nor in the particle's eccentricity $e$. 
The methodology used is based on a semi-analytical model that consists on calculating the averaged resonant disturbing function numerically, assuming for this that in the resonant scale of time all the orbital elements of the particle are constant. 
In order to obtain the secular evolution inside the MMR, we make use of the adiabatic invariance principle, assuming a zero-amplitude resonant libration. We construct two-dimensional surfaces (called $\mathcal{H}$ surfaces) in the three-dimensional space $(\sigma, e, \varpi)$ that allow us to predict the secular evolution of these three variables.
The 2:1 MMR is used as example to show some results. We found four apsidal corotation resonance (ACR) families, two symmetric and two asymmetric. One of the symmetric families exists for almost any $e_p$ value. The other one for $e_p>0.3$ and the asymmetric ones for $e_p>0.44$. We corroborate the secular variations in $e$ and $\varpi$ predicted by the model through numerical integrations even when the initial conditions are displaced from those ACR.
Some peculiar examples are presented for the 3:1 and 3:2 MMR showing large excursions in eccentricity. As an application, the Planet 9 is investigated as a possible responsible of high eccentric distant TNOs.

\end{abstract}


%%%%%%%%%%%%%%%%%%%%%%%%%%%%%%%%%%%%%%%%%%%%%%%%%%

%%%%%%%%%%%%%%%%% BODY OF PAPER %%%%%%%%%%%%%%%%%%

\section{Introduction}
\label{sec:intro}

% FIGURA 1
\begin{figure}
  \includegraphics[width=\linewidth]{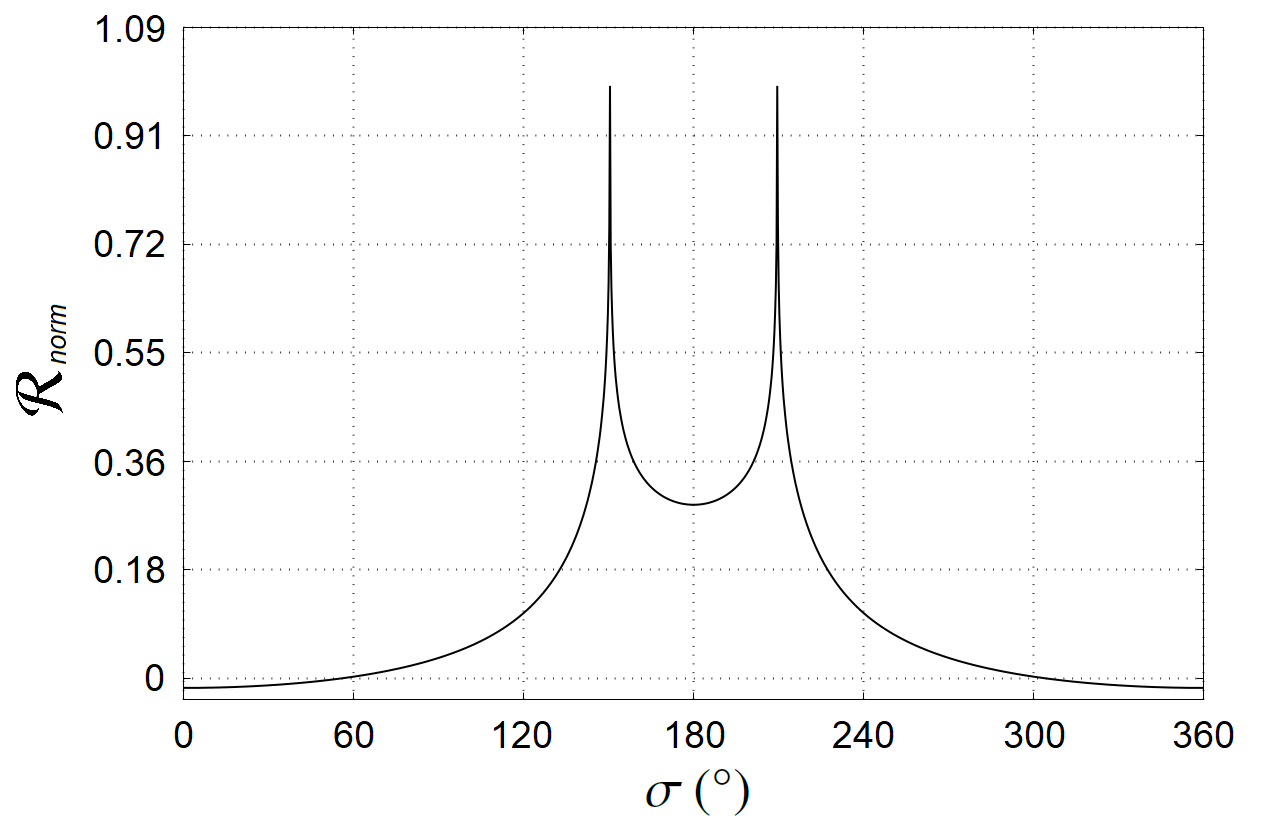}
  \caption{Normalised averaged disturbing function for the 2:1 MMR with $\varpi=\varpi_p=0$°, $e = 0.73$ and $e_p = 0.01$ (quasi-circular case).}
  \label{fig:rsigma}
\end{figure}

Resonant motions, in the celestial mechanics context, have been object of intense study for many decades due to the diverse dynamical evolution they can produce. The initial efforts on analytical theories developments resulted in the first and second fundamental resonant models \citep{1966AJ.....71..657G,1983CeMec..30..197H} which were applicable to a particle in mean motion resonance (MMR) with a circular perturber. They had the advantage of being fully integrable as they are hamiltonian systems of one degree of freedom. For non-zero low eccentricity of the perturber $e_p$ and of the particle $e$, several analytical expansions in $e_p$ and $e$ for the disturbing function $\mathcal{R}$ exist. Some classical examples of these are in \citet{1982AJ.....87..577W,1985Icar...63..272W}. Naturally, these expansions are valid for small eccentricities which constitutes their main limitation. As an alternative, there are other type of expansions called asymmetric expansions \citep{1987A&A...183..397F} that are implemented around a general $e>0$ value. If the variation of $e$ is too large, another expansion can be done to continue studying the secular evolution of the particle. There are some other works, for instance \citet{1993CeMDA..57...99M,1995Icar..114...33M}, that only expanded in $e_p$, allowing an application to any high value of $e$, with the restriction $e>e_p$. \citet{2006CosRe..44..440S} also used only a Laplacian expansion in $e_p$ but without any restriction for $e$. That method was based on a double numerical average that allow to study the secular evolution of asteroids inside the 3:1 MMR with Jupiter. 
In the restricted case there are also works that did not use analytical expansions at all, as for example in \citet{1989A&A...213..436Y} where the long-term changes of asteroid's eccentricities in several MMR with Jupiter were calculated. There, it was considered all Jupiter´s parameters fixed except for its longitude of perihelion, assumed to change linearly with time. With a similar methodology, \citet{1996Icar..120..358B} presented various phase portraits in the $(e,\varpi)$ domain for some MMR. More recently \citet{2017A&A...605A..23P} studied the elliptic restricted problem also without analytical expansions and developed a similar approach as we will present here. Our work is in the same line of disregarding analytical expansions in order to have a valid method for extreme variations of $e$ and arbitrary values of $e_p$. We are also interested in the long-term evolution of the libration centre. Variations of the resonant libration centre have been already observed in systems with mutual inclination, for example, in \citet{2006Icar..181..205G}. In the present work we show the same phenomenon can occur for coplanar high-eccentricity systems. 

% FIGURA 2
\begin{figure}
  \includegraphics[width=\linewidth]{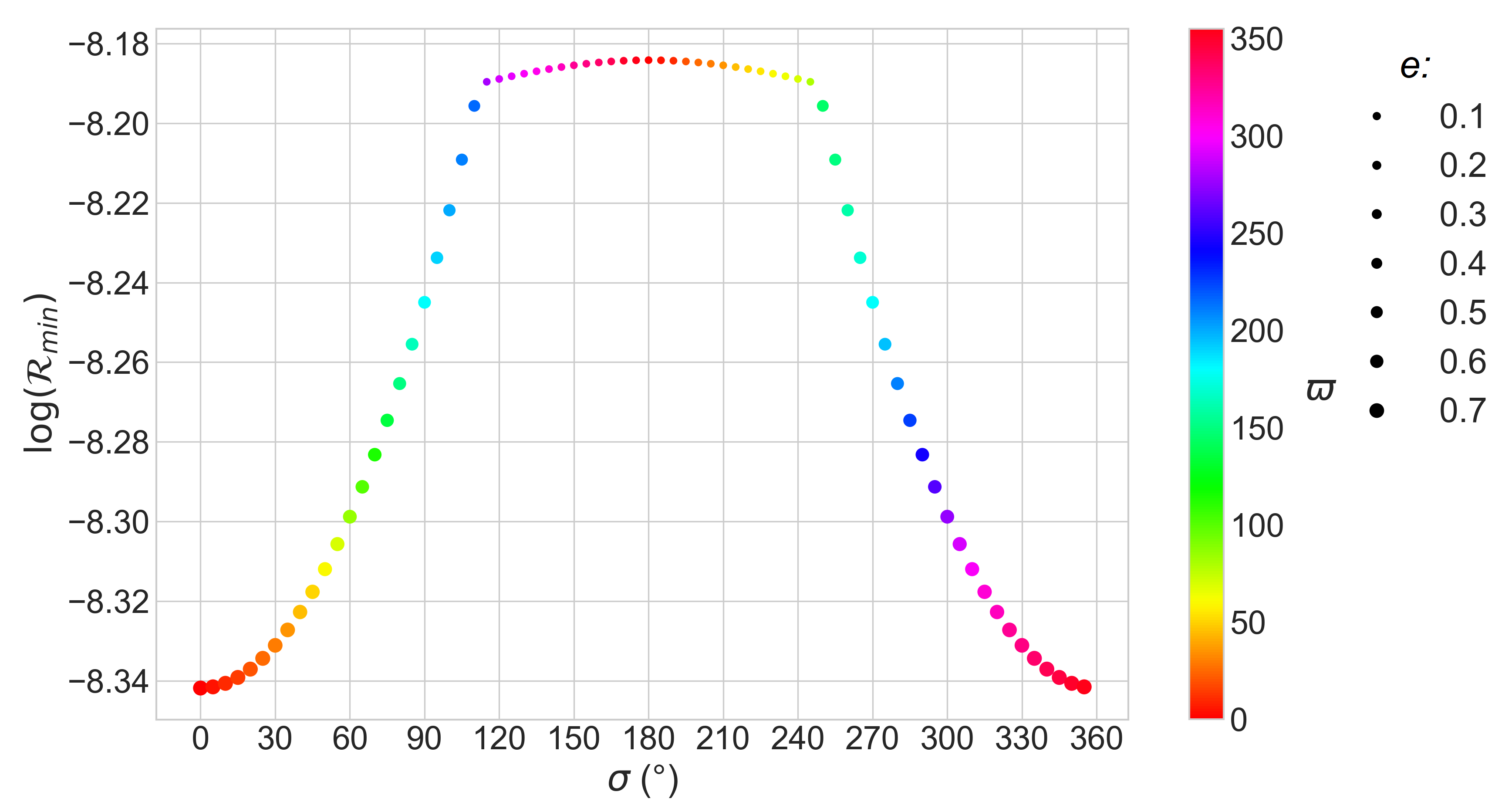}
  \caption{$\min{\{\mathcal{R}(e,\varpi)\}}$ vs $\sigma$ for the 2:1 MMR with $e_p = 0.01$.}
  \label{fig:2-1_e2=0.01_Rmin_sigma}
\end{figure}

% FIGURA 3
\begin{figure}
  \includegraphics[width=\linewidth]{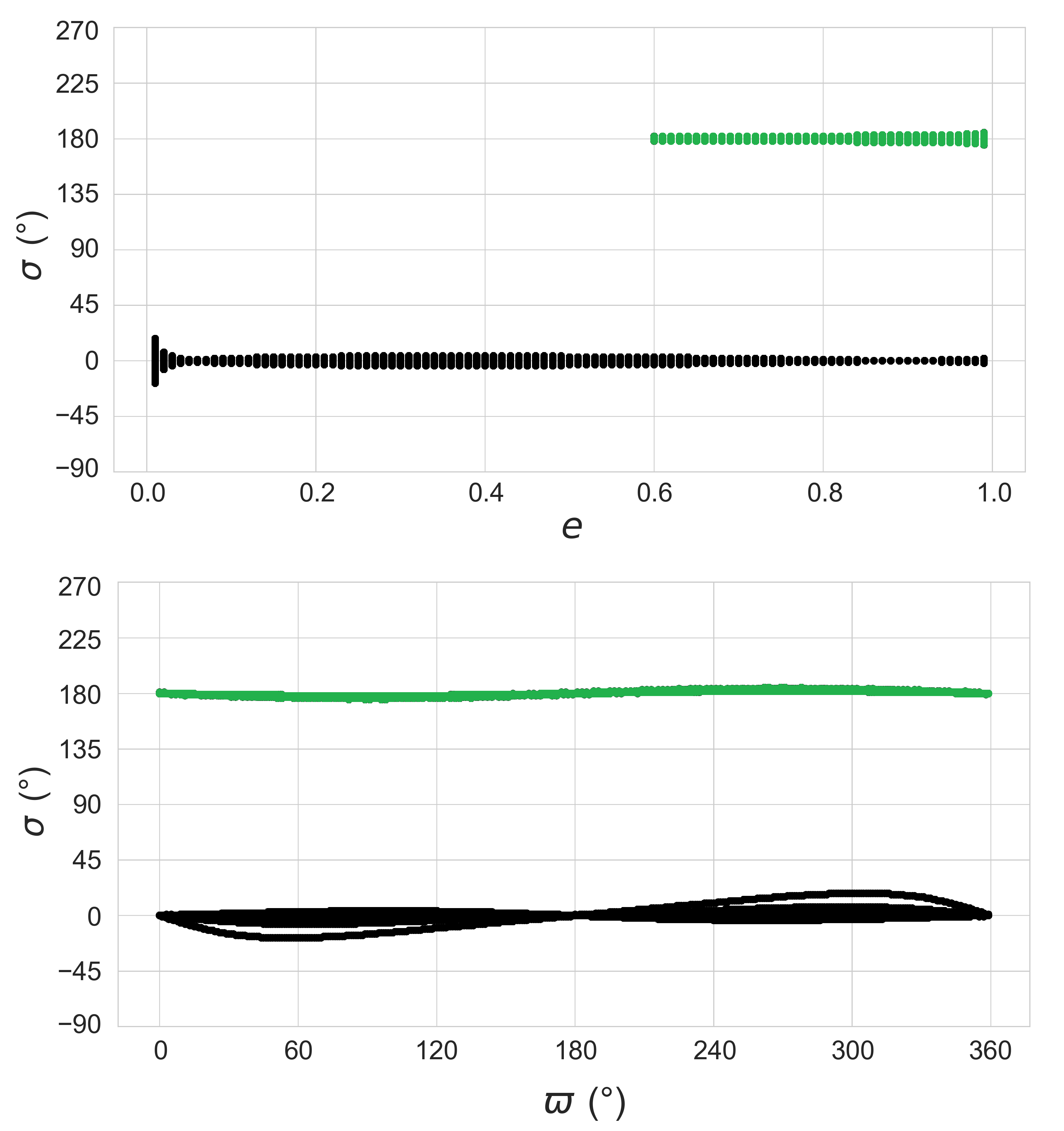}
  \caption{Equilibrium points calculated with equation \ref{eq:dR_ds} for the 2:1 MMR with $e_p = 0.01$.}
  \label{fig:2-1_e2=0.01_sigma_vs_e1}
\end{figure}

With respect to pure numerical techniques, there are various works with different methodologies, for example, \citet{2018CeMDA.130...41A} analysed several MMRs through stability maps using a chaos indicator, \citet{2003ApJ...596.1332H} studied the problem via searching resonant periodic orbits with the differential continuation method, \citet{Celletti2002} solved numerically the differential equations looking for stable mirror configurations, etc. These are interesting works with the disadvantage that sometimes are computing consuming and could be difficult to reveal global dynamical features.

All these studies are somehow complementary and contribute to the understanding of different dynamical aspects of the MMR. In the present work, we extend the approach of \citet{2020CeMDA.132....9G} studying the secular evolution of the restricted coplanar resonant case for any eccentricity of both bodies and find the long-term evolution of the equilibrium points in the space ($e,\varpi,\sigma$). Our model is validated through the comparison with numerical integrations of the full equations of motion obtaining very good agreement. 
A similar approach was developed in \citet{Li_2021}, where they studied the 1:1 MMR. The only disadvantage in our method is that the results can be erroneous for low $e$ and $e_p$ in first order MMRs. This occurs because in those cases $\dot\varpi$ is too high and, as we will show, this invalidates two hypothesis we will use to develop our model. 
One consequence of this is that we do not reproduce the known law of structure \citep{1988ltdb.conf..245F} that relates $e$ with $a$ when $e\rightarrow0$.

Despite no having an immediate innovative application in the Solar System because of the planet's low eccentricities, it could be applied to some extrasolar systems with two planets in MMR being one of them much more massive than the other or to exoasteroids/exocomets in MMR with an eccentric exoplanet. This last application is too far away from being able to be contrasted with observations at present for obvious reasons. In this work we apply it to the hypothetically Planet 9 and show that it could be a partial responsible of some orbital characteristics in distant TNOs.

\section{Theoretical framework and methodology}
\label{sec:Theory}

\subsection{Semi-analytical model}
\label{sec:model}

% FIGURA 4
\begin{figure*}
  \includegraphics[width=0.9\textwidth]{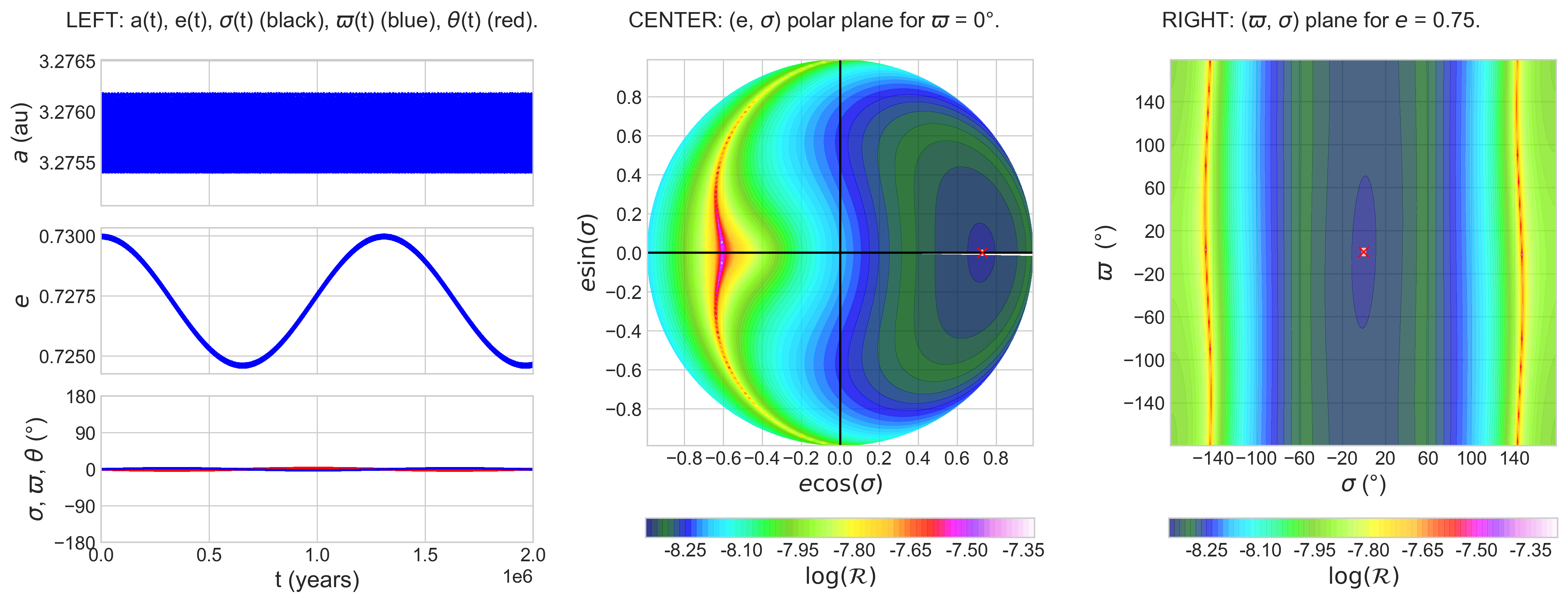}
  \caption{2:1 MMR with $e_p = 0.01$. 2 Myrs numerical integration and the $\mathcal{R}(\sigma,e)$ and $\mathcal{R}(\sigma,\varpi)$ contour maps are shown.}
  \label{fig:2-1_e2=0.01_int_Re1s1_Rw1s1}
\end{figure*}

% FIGURA 5
\begin{figure*}
  \includegraphics[width=0.9\textwidth]{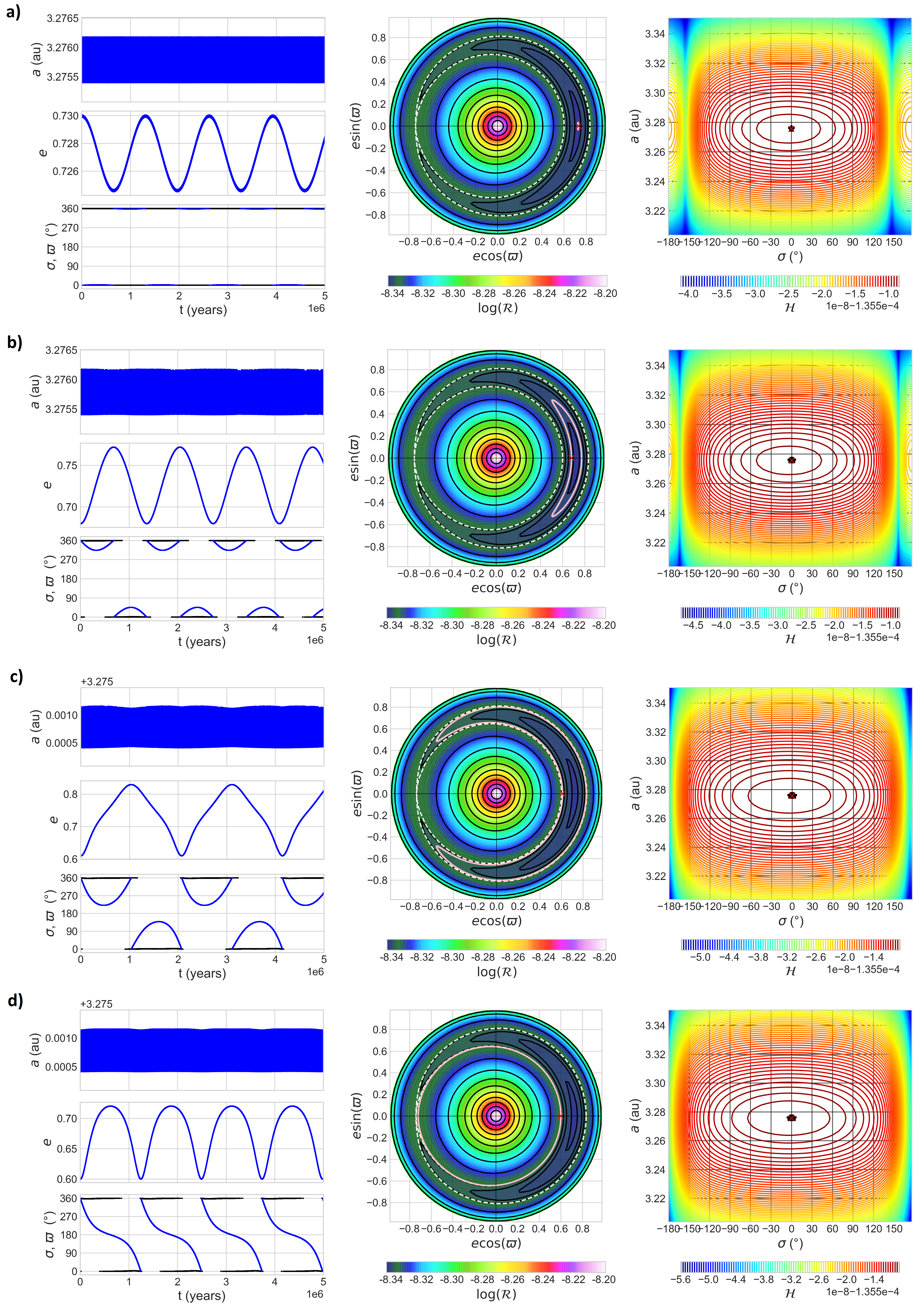}
  \caption{2:1 MMR with $e_p = 0.01$ \textit{LEFT}: $a(t)$, $e(t)$, $\sigma(t)$ (black) and $\varpi(t)$ (blue) from the numerical integrations. \textit{CENTRE}: $\mathcal{R}(e,\varpi)$ map for $\sigma=0$° Vs. numerical integration in pink. The white-dashed curve corresponds to the separatrix. \textit{RIGHT}: $\mathcal{H}(a, \sigma)$ contour curves for the $(e_i,\varpi_i)$ pair Vs. numerical integration in black. From top to bottom the difference is in the $e_i$: a) $0.73$. b) $0.68$. c) $0.61$. d) $0.60$. }
  \label{fig:2-1_e2=0.01_int_Re1w1_Ha1s1}
\end{figure*}

In the rest of the article every orbital element without sub-index refers to the particle whereas sub-index "p" refers to the perturbing planet (massive body). The "s" index is reserved for the star. 
 The method we will devise is valid for arbitrary resonances but
in this work we focus on internal resonances, i.e., the semi-major axes satisfy $a<a_p$ always.

As we are assuming a coplanar configuration, only three orbital elements are relevant which are $a$, the eccentricity $e$ and the longitude of the pericenter $\varpi$ (formally would be $\Delta\varpi=\varpi-\varpi_p$ but we will assume $\varpi_p=0$). Besides these, due to the resonant hypothesis, the critical angle $\sigma$ is also a relevant parameter, which is defined as follows:
\begin{equation}
    \sigma = k\lambda - k_p\lambda_p + (k_p-k)\varpi
    \label{eq:sigma}
\end{equation}

where $\lambda$, $\lambda_p$ are the mean longitudes and $k,k_p$ are positive integers. The planet's orbital elements are evidently fixed since we are in the restricted case.

Following, for example, \citet{2002aste.book..379N} or \citet{2016CeMDA.126..369S} the semi-secular hamiltonian obtained eliminating the short-period terms depending on $\lambda$ or $\lambda_p$, but no $\sigma$, is:
\begin{equation}
    \mathcal{H}(a, e, \varpi, \sigma) = -\frac{\mu}{2a}-n_p\frac{k_p}{k}\sqrt{\mu a} - \mathcal{R}(a, e, \varpi, \sigma)
    \label{eq:hamiltonian}
\end{equation}

where $\mu = Gm_s$ and $n_p$ is the planet's mean motion.

The hamiltonian of equation \ref{eq:hamiltonian} has three terms, the first one is the Keplerian term, the second one corresponds to the expanded phase space term in order to have an autonomous $\mathcal{H}$ and the third term is the averaged disturbing function. To calculate it we follow the idea given by \citet{1968AJ.....73...99S} with the approach used by \citet{2006Icar..184...29G, 2019Icar..317..121G, 2020CeMDA.132....9G} where the resonant disturbing function is calculated by an averaging method in $\lambda_p$, considering that $e$ and $\varpi$ are fixed. This results in a great simplification since the problem becomes of one degree of freedom.

Using the canonical variables $\Sigma = \sqrt{\mu a}/k$ and $\sigma$, the equilibrium points have to satisfy the following conditions:
\begin{equation}
    \frac{\partial\mathcal{H}}{\partial\Sigma}=0 ; \qquad \frac{\partial\mathcal{H}}{\partial\sigma}=0
    \label{eq:canon_eqs}
\end{equation}

The first one gives the nominal semi-major axis:
\begin{equation}
    a=\frac{a_p}{(1+m_p/m_s)^{1/3}}\left(\frac{k}{k_p}\right)^{2/3}\equiv a_0
    \label{eq:a_nominal}
\end{equation}
where it was neglected the contribution of $\frac{\partial\mathcal{R}}{\partial\Sigma}$ as it is proportional to $m_p<<m_s$. This simplification does not change significantly the value of $a_0$.

We can rewrite the second condition in \ref{eq:canon_eqs}, which gives $\sigma_0$, known as the equilibrium centre of libration:
\begin{equation}
    \frac{d\mathcal{R}}{d\sigma}=0
    \label{eq:dR_ds}
\end{equation}

Strictly speaking this is correct but we are interested in the stable equilibrium points. They occur when $\sigma$ minimises (at least locally) the function $\mathcal{R}$. If we consider a constant semi-major axis for the particle, then is totally equivalent finding $\mathcal{R}$ minimums to finding $\mathcal{H}$ maximums. From here on, we will refer to stable equilibrium points just as equilibrium points. We are going to disregard those equilibrium points obtained when an encounter occurs. The criteria used to detect an encounter is given by the next inequality:
\begin{equation}
    \Delta < \eta R_{H} = \eta a_p\left(\frac{m_p}{3m_s}\right)^{1/3} 
    \label{eq:enc}
\end{equation}
where $R_{H}$ is the Hill's radius, $\Delta$ is the minimal distance between the bodies and $\eta$ is a tolerance factor with typical values between 2 and 4. We use $\eta=3$ for the examples presented in this work.

In the Fig. \ref{fig:rsigma} there is a $\mathcal{R}(\sigma)$ example in the 2:1 resonance, where $\varpi=\varpi_p=0$°, $e = 0.73$ and $e_p = 0.01$. Is easy to see that there are two equilibrium points, one at $\sigma=0$° and the other at $\sigma=180$°. The second one is surrounded by two high spikes due to encounters between the bodies. In general, these equilibrium points found with this method are always stable in relatively short scales of time ($t \sim$ 20 -- 200 $P$ being $P$ the orbital period) when numerical integrations are carried out. However, if we check their stability at higher scales of time ($t \sim 2\times 10^4$ -- $2\times 10^6$ $P$) some points remain stable whereas others not.

To overcome this issue and find the long-term stable points, we use the adiabatic invariance principle which has been applied in MMR dynamics at least since \citet{1976rapp.rept...15P}. In that work the author applied it to study the capture process in MMR due to tidal interactions. A recent example where the principle is applied to study secular evolutions in MMR can be found in \citet{2017_Batygin_Morbidelli}. This principle states that the adiabatic invariant of the dynamics $J$ remains constant as long as the resonant libration periods are much more shorter than the secular periods \citep{Henrard1993}. The definition of this quantity is as follows:
\begin{equation}
    J=\oint \Sigma d\sigma
    \label{eq:J}
\end{equation}

For the sake of simplicity we are going to work in a negligible resonant amplitude of libration regime ($J=0$), which means that $a=a_0$ and $\sigma=\sigma_0$ are essentially constants in resonant time-scales, with zero-amplitude resonant librations. This approximation was used for example by \citet{1985CeMec..36...47K} and subsequent works. In secular time-scales $a$ will continue to be constant (because of the commensurability between orbital periods) but the centre of resonant libration could slowly change.

In practice this means that some of the orbital elements are slow varying and could be treated as constants to do the averaging of $\mathcal{R}$, which is carried out in a shorter time-scale. Therefore, when the secular evolution of $e$ and $\varpi$ is much slower than the resonant libration periods of $\sigma$, we can apply the adiabatic invariance principle and do the averaging for all the possible $(e, \varpi)$ pairs. This allow us to construct contour maps of the type $\mathcal{R}(\sigma,e) = C$, $\mathcal{R}(\sigma,\varpi) = C$ or $\mathcal{R}(e,\varpi) = C$ being $C$ constant (from now on, $C$ will always refer to an arbitrary constant) that will help with the understanding of the secular evolution. In each one of these maps, the missing variable has to be at least constant enough so they remain valid when compared with long-term numerical integration. The allowed variation to ensure this map's validity will depend on each particular case.

Another more general way of studying the evolution and very useful in more complicated cases is searching for all the equilibrium points in the $(\sigma,e,\varpi)$ space using equation \ref{eq:dR_ds}. Then, using the fact that $\mathcal{H}$ is a constant of motion and the adiabatic invariance principle, a kind of three dimensional contours curves in the mentioned space can be constructed that will predict the complete secular evolution of $\sigma$, $e$ and $\varpi$. We will call them hamiltonian 3D maps or simply $\mathcal{H}$ surfaces.

\subsection{Methodology}
\label{sec:Methodolgy}

For all the cases, is assumed a central star of mass $m_s=1 M_\odot$ and a unique planet with $a_p$ = 5.2 au, i.e., same as Jupiter, but with 10\% of its mass.
Without losing generality, both, the planet's longitude of the pericenter and the mean anomaly were set to zero, so, $\varpi_p=M_p=0$. On the other hand, $\varpi$ could vary and the particle's mean anomaly $M$ is defined from the critical angle $\sigma$ through the following relation (deduced from equation \ref{eq:sigma}):
\begin{equation}
    M =  \frac{\sigma - k_p\varpi}{k}
    \label{eq:M1}
\end{equation}

This equation seems to tell us that only one mean anomaly could produce the deep resonant behaviour. In fact this is true, but only for $k=1$. If $k=2$, we could increase $\sigma$ by 2$\pi$ and we would obtain two different mean anomalies $M_1$ and $M_2$ such that $|M_1-M_2|=\pi$. In general, for any arbitrary $k$ value, there will be $k$ different mean anomalies that will satisfy the resonant condition.

We introduce the angle $\theta$ which is defined as follow:
\begin{equation}
    \theta = k\lambda - k_p\lambda_p + (k_p-k)\varpi_p
    \label{eq:theta}
\end{equation}

Since we are supposing $\varpi_p=0$, then $\theta= k\lambda - k_p\lambda_p$, i.e., just the linear combination of mean longitudes. This angle is simply another resonant argument which could be more relevant to analyse the resonant motion in those cases where $e_p>e$. 

In order to be rather exhaustive, given a resonance $k_p:k$, we explore the $(\sigma,e,\varpi)$ space for $e_p$ in the range (0.01, 0.85). We analysed more in detail the cases with the following specific values: $e_p = (0.01; 0.3; 0.5)$. 
Once $k$, $k_p$ and $e_p$ are fixed, the space $(\sigma,e,\varpi)$ can be explored in different ways, detailed in the following sections. In the section \ref{sec:Rs1e1_Rs1w1_maps} is described a first method to explore and search for secular equilibrium points, i.e., points where all the variables are static for long periods of time. In fact, these points have been widely studied in the planetary case \citep{2003MNRAS.341..760B,2004MNRAS.350.1495Z,2008MNRAS.387..747M} and are called apsidal corotation resonances (ACR from here on). In the section \ref{sec:Re1w1_maps} is described a method to predict the secular behaviour of $e$ and $\varpi$ in those cases where the centre of resonant libration is constant, despite $e$ and $\varpi$ changing in time. Finally, in the section \ref{sec:H_maps} is presented the technique developed to study the secular evolution in the more general case of a variable centre of libration.

\subsubsection{$\mathcal{R}(\sigma,e)$ and $\mathcal{R}(\sigma,\varpi)$ maps}
\label{sec:Rs1e1_Rs1w1_maps}

$\mathcal{R}(\sigma,e)$ maps can be made assuming some fixed value for $\varpi$. Alternatively, if $e$ is assumed fixed, $\mathcal{R}(\sigma,\varpi)$ maps can be constructed. These plots are just the contour curves of $\mathcal{R}$ and are useful to find ACRs as long as a minimum in the $(\sigma,e)$ space matches with a minimum in the $(\sigma,\varpi)$ space, resulting in a minimum in the entire $(\sigma,e,\varpi)$ space. A faster approach could be to sweep $\sigma$ from $0$ to $360$° obtaining in each step the $\min{\{\mathcal{R}(e,\varpi)\}}$. After this, a curve as the one in the Fig. \ref{fig:2-1_e2=0.01_Rmin_sigma} can be plotted, revealing for example the absolute $\min{\{\mathcal{R}(\sigma, e, \varpi)\}}$ or other local minimums. 
We will show that if a numerical integration is set in the absolute minimum of $\mathcal{R}$, the orbit freezes and does not show any change over time.

A problem with this first method arises when a variation of $e$ or $\varpi$ implies a topological change in one of this two-dimensional $\mathcal{R}$ contour curves. These variations could exist depending if $\frac{\partial\mathcal{R}}{\partial\varpi}\neq0$ and/or $\frac{\partial\mathcal{R}}{\partial e}\neq0$. If both are zero, no variation will exist. But, for example, if a variation in $e$ implies that the map $\mathcal{R}(\sigma,\varpi)$ changes topologically, then it means that is no longer always valid, so, numerical integrations cannot be contrasted with this map in all the numerical integration time interval.

\subsubsection{$\mathcal{R}(e,\varpi)$ maps}
\label{sec:Re1w1_maps}

If we assume that the centre of libration is fixed in the secular time-scale for any $e$ and $\varpi$ values, the numerical integrations should follow the contour curves of $\mathcal{R}(e,\varpi)$. To test this hypothesis, a double sweep in $e$ and $\varpi$ can be made, registering the $\sigma$ equilibrium values in each step. This allows to plot something like the graphic shown in Fig. \ref{fig:2-1_e2=0.01_sigma_vs_e1}. As can be seen there, there are always equilibrium points very close to $\sigma = 0$°. This means that no topological change (or at least near $\sigma = 0$°) exists for $\mathcal{R}$. The equilibrium points near $\sigma=180$° (coloured in green) are other family of equilibrium points which appear at high $e$ values. Thoroughly speaking, the mentioned sweep to check if the centre of libration is fixed, would be required only for the range of variation of $(e,\varpi)$ that the secular evolution would induce. Nevertheless, is important to have in mind if the entire $\mathcal{R}(e,\varpi)$ map is valid and describes the correct dynamics or if only is valid a sub-region of it. 

For $e\rightarrow0$ there is a small dispersion from $\sigma=0$°. In that zone, the model's validity is compromised for first order MMR, as we mentioned, because $\dot\varpi$ is too high invalidating the calculation of $\mathcal{R}$ assuming that $(a, e, \varpi)$ are fixed during the considered averaging period. Besides, the adiabatic invariant principle cannot be applied since $\varpi$ circulates too fast and its frequency could be comparable with $\sigma$ libration frequency. This is associated with the undefined intrinsic characteristic of $\varpi$ when $e\rightarrow0$. In those low $e$ equilibrium points the resonant strength is usually low \citep{2019Icar..317..121G}.

As in the previous section, we will show that if a numerical integration is set in the absolute minimum of $\mathcal{R}$, every orbital element will be constant. The interesting result comes out when the initial conditions are not in the ACR point, for example, changing the initial $e$ value, and finding out that the secular evolution of $e(t)$ and $\varpi(t)$ follows almost exactly the contour curves of $\mathcal{R}(e,\varpi)$. As we mentioned, this would be the case if the asteroid is locked in the same resonant libration centre with zero-amplitude libration ($\dot{\sigma}\equiv0$ and $J=0$).

\subsubsection{$\mathcal{H}$ surfaces}
\label{sec:H_maps}

In the most general case that the secular evolution occurs for the three variables $e$, $\varpi$ and $\sigma$, i.e., the libration centre varies, a three-dimensional representation is needed. This representation is a 2D surface in the 3D space $(\sigma,e,\varpi)$.

This $\mathcal{H}$ surface is conformed by all the resonant equilibrium points which can be visualised when they are plot altogether in the $(\sigma, e, \varpi)$ space. This surface contains the contour curves given by $\mathcal{H}=C$, which gives the possible secular dynamic trajectories of the system. Once the initial $(e, \varpi)$ pair is defined and $\sigma$ satisfies equation \ref{eq:dR_ds} (in order to be in the surface), the secular evolution of $(\sigma, e, \varpi)$ is given by one of these curves (the one defined by the initial $\mathcal{H}$). Formally, it is required not only the verification of equation \ref{eq:dR_ds} but also that this point is a minimum of $\mathcal{R}$ or a maximum of $\mathcal{H}$. 

Being exactly in those resonant equilibrium points guarantees that we are in a $J=0$ framework, i.e., in the zero-amplitude resonant libration hypothesis. This will be the reason why the secular evolution could be predicted by these three-dimensional contour curves. In this situation we could interpret that the secular evolution modifies the resonant centre of libration and the asteroid "follows" it always maintaining a zero-amplitude libration, as long as this centre does not change too fast.

If $\sigma$ changes too fast, the adiabatic invariance principle could fail and $J$ could increase, causing a non-zero amplitude of resonant libration. If this amplitude is positive but small, these maps still would represent the evolution good enough where a small deviation from the $\mathcal{H}$ surface can be observed. If $J$ increases too much, the comparison between the numerical integration and the map could become rapidly uncorrelated.

% FIGURA 6
\begin{figure*}
  \includegraphics[width=0.99\textwidth]{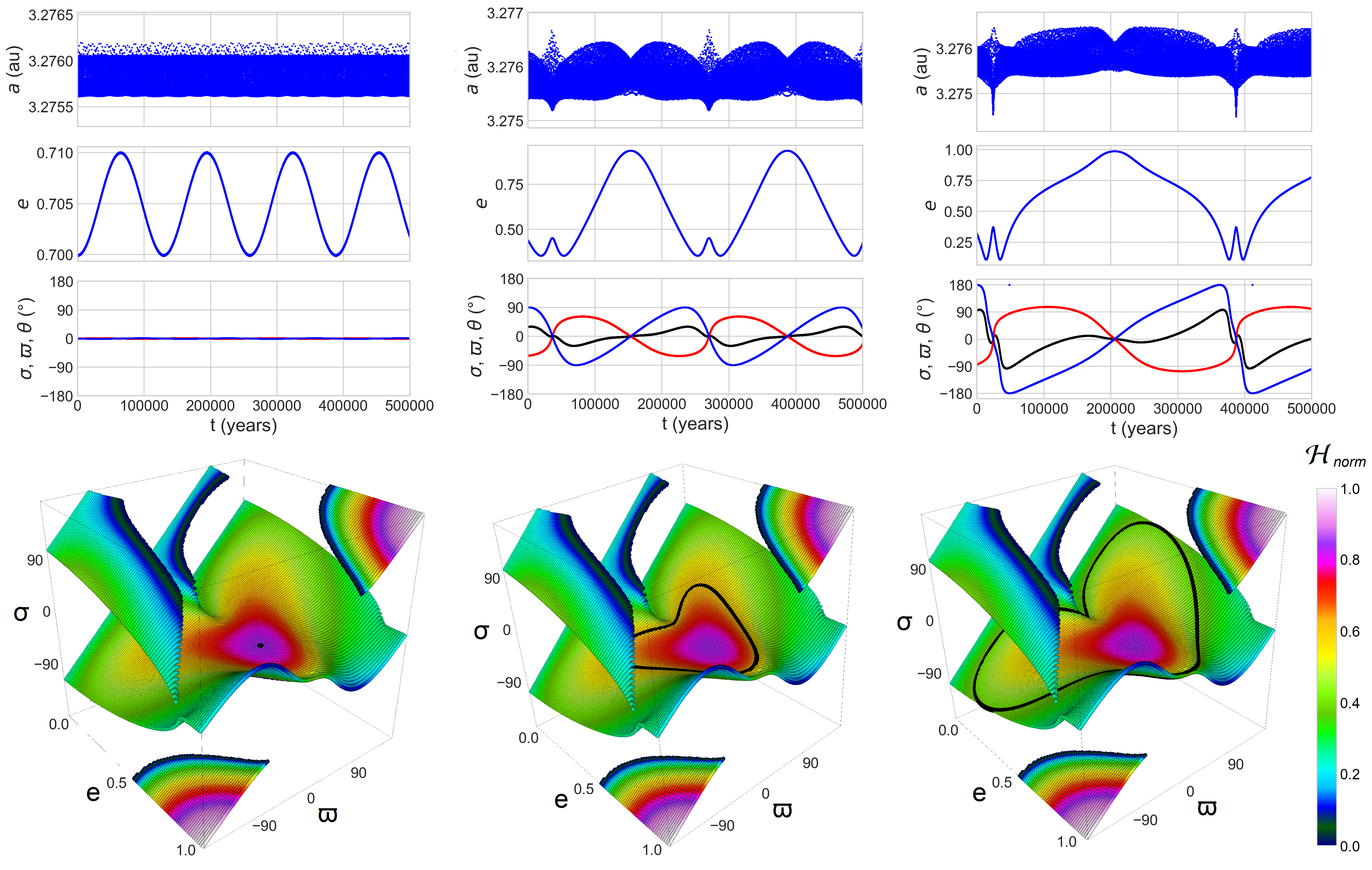}
  \caption{2:1 MMR with $e_p=0.3$. \textit{TOP:} Three examples of numerical integrations. In blue are $a$, $e$ and $\varpi$. In black is $\sigma$ whereas in red is $\theta$. \textit{BOTTOM:} Comparison between the $\mathcal{H}$ surfaces and each of the numerical integrations (black curves). The ($\sigma_i$, $e_i$, $\varpi_i$) are the following $\rightarrow$ \textit{LEFT:} (0°, 0.7, 0°). \textit{CENTRE:} (29°, 0.44, 90°). \textit{RIGHT:} (95°, 0.32, 180°).}
  \label{fig:2-1_3int_vs_3DHmaps}
\end{figure*}

\section{Results}
\label{sec:results}

\subsection{The 2:1 MMR}
\label{sec:2-1}

In this section are presented some examples for $k=1$ and $k_p=2$. In the appendix are presented two short examples for the 3:1 and 3:2 MMR. As we have stated before, in this work we only consider internal resonances but this method can be applied to exterior resonances and even for the 1:1 MMR. From here on we are going to use sub-index "i" referring to "initial" and sub-index "0" for the ACR points.

\subsubsection{Quasi-circular planet ($e_p=0.01$)}
\label{sec_ep=0.01}

In the Fig. \ref{fig:2-1_e2=0.01_int_Re1s1_Rw1s1} we show the $\mathcal{R}(\sigma,e)$ and $\mathcal{R}(\sigma,\varpi)$ maps for $e_p=0.01$ compared with a numerical integration of the exact equations of motion which initial conditions are those of the ACR point, i.e., $(\sigma_i, e_i, \varpi_i)=(\sigma_0, e_0, \varpi_0) = (0^{\circ},0.73, 0^{\circ})$. Over the maps is a red cross indicating the initial condition and in pink is the numerical integration itself, which in this case is barely visible because, as expected, no secular variations in the orbital elements occur. The highest values of $\mathcal{R}$ (from red to white colours) corresponds to the encounter zone. The corotational solution for $e_p\simeq0$ in this MMR has been well known since \cite{1993CeMDA..55...25F}.

% FIGURA 7
\begin{figure*}
  \includegraphics[width=\textwidth]{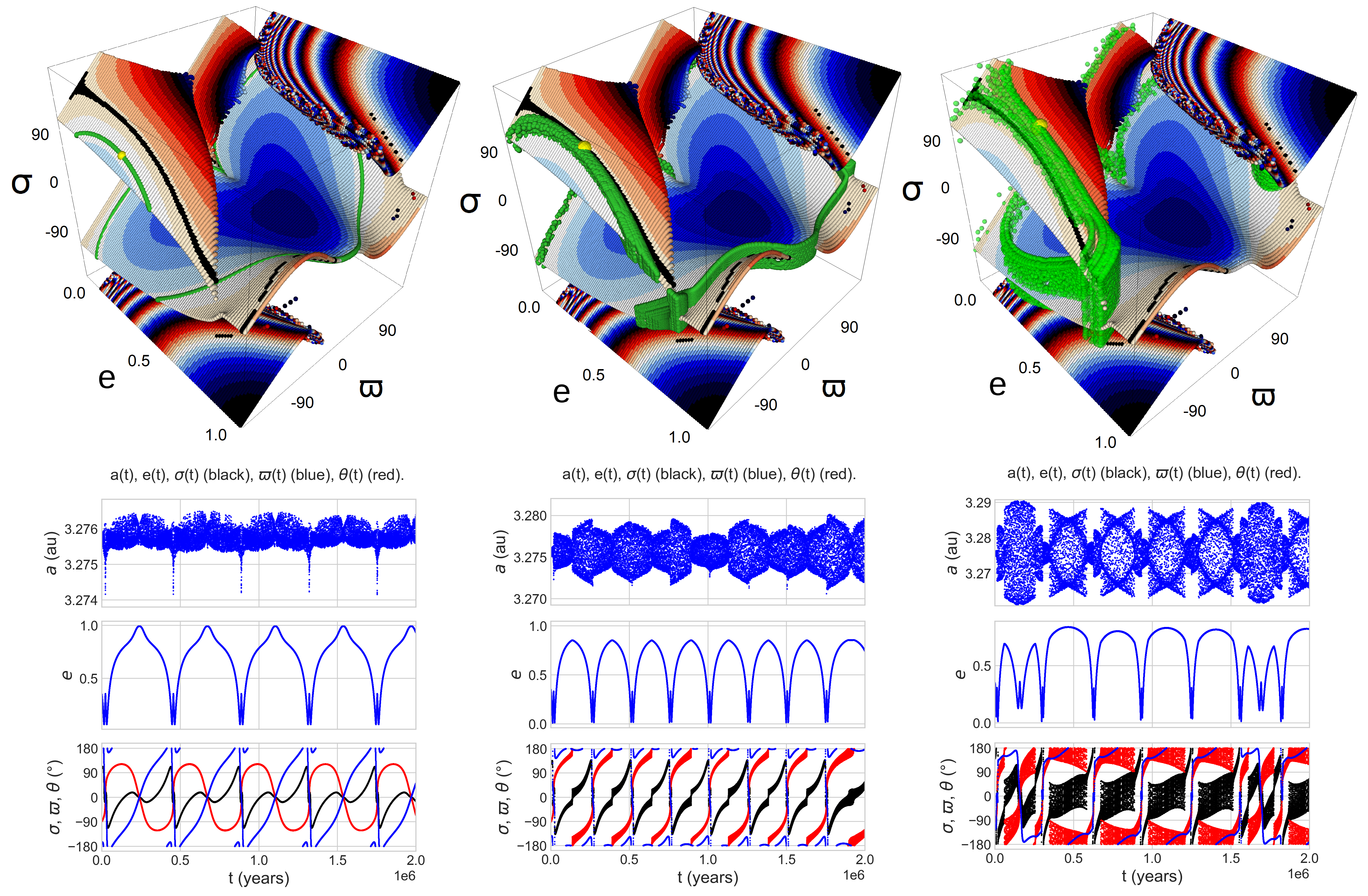}
  \caption{2:1 MMR with $e_p=0.3$. Three numerical integrations (in green) are compared with the $\mathcal{H}$ surfaces to illustrate the breakdown of $J=0$ hypothesis. The black curve is the separatrix. $(\sigma_i,e_i,\varpi_i)=$ \textit{LEFT}: (106°, 0.35, 195°). \textit{CENTRE}: (123°, 0.35, 215°). \textit{RIGHT}: (140°, 0.35, 235°). No filtration of equilibrium points has been done here to construct the $\mathcal{H}$ surface.}
  \label{fig:2-1_e2=0.3_separatriz}
\end{figure*}

In order to study what happens around the ACR we are going to change $e_i$ from $e_0$ but maintaining $\varpi_i=\varpi_0$ and $\sigma_i=\sigma_0$. The Fig. \ref{fig:2-1_e2=0.01_int_Re1w1_Ha1s1} shows what happens if this is done. We have plotted the numerical integrations in the time domain, the $\mathcal{R}(e, \varpi)$ maps compared with the numerical integration in pink and the contour curves of $\mathcal{H}(a, \sigma)$ calculated from equation \ref{eq:hamiltonian} for the initial $(e_i, \varpi_i)$ of the numerical integration, which is in black.

Note how the secular evolution of the eccentricity and the longitude of the pericenter matches pretty well with the $\mathcal{R}(e, \varpi)$ map's contour curves. Moreover, the change in behaviour of $\varpi$ between oscillating and circulating also agrees with the predicted by these contour curves, occurring in this case for $e\simeq0.6$. From these results we can conclude that a small variation in $e_i$ from $e_0$ can produce large-amplitude oscillations in $\varpi$. The variations induced in $e$ itself are also significant. Let be $\Delta e = e_{max} - e_{min}$. In the $\varpi$ libration regime, it is satisfied $\Delta e \simeq 2(e_0 - e_i)$. When $\varpi$ circulates, the variations in $e$ are smaller compared to the example where the particle was in the edge of the $\varpi$ libration regime (cases c vs d in fig. \ref{fig:2-1_e2=0.01_int_Re1w1_Ha1s1}).

At this point an important remark regarding the hamiltonian must be done. When $e$ and/or $\varpi$ changes, the global topology of the $\mathcal{H}(a,\sigma)$ contour levels could change. If we compare, for example the a) map with the d) one, is evident how the $\sigma=180$° libration centre disappears. This is in accordance with the graphic in Fig. \ref{fig:2-1_e2=0.01_sigma_vs_e1}. In general some new families of equilibrium points could (dis)appear, and also the libration centre value (i.e. $\sigma$ itself) of some of these families could change, as we will show. But in this case there is always an equilibrium point in $\sigma = 0°$, regardless of $e$ and $\varpi$ values. Therefore, the analysis via the $\mathcal{R}(e,\varpi)$ contour levels in the Fig. \ref{fig:2-1_e2=0.01_int_Re1w1_Ha1s1} is valid and sufficient, as we are supposing that the adiabatic invariance principle is correct. As we mentioned, there will be some cases that large variations of $e$ and/or $\varpi$ would imply great modifications in the centre of resonant librations. The next case is an example of this.

% FIGURA 8
\begin{figure*}
  \includegraphics[width=\textwidth]{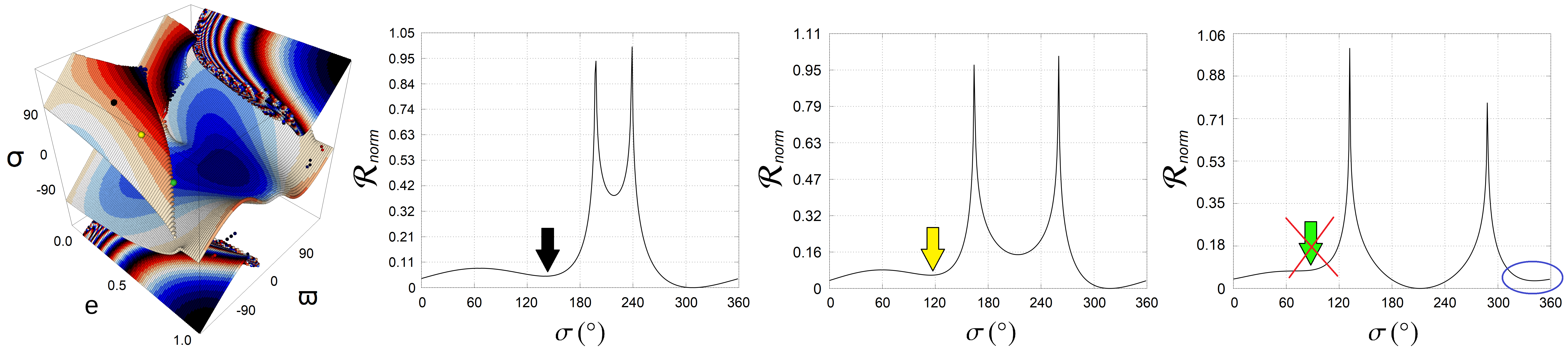}
  \caption{2:1 MMR with $e_p=0.3$. \textit{LEFT:} $\mathcal{H}$ surface with three points in the open curve zone marked with different colours but laying in the same curve. \textit{RIGHT:} The 3 functions $\mathcal{R}(\sigma)$ for the 3 coloured points. $(e,\varpi)_{black}=$ (0.35, 235) ;  $(e,\varpi)_{yellow}=$ (0.55, 226) ;  $(e,\varpi)_{green}=$ (0.75, 220).}
  \label{fig:2-1_e2=0.3_3ptos}
\end{figure*}

\subsubsection{High-eccentricity planet ($e_p=0.3$)}
\label{sec_ep=0.3}

Following the procedure described in \ref{sec:Rs1e1_Rs1w1_maps}, a main ACR in $(\sigma_0,e_0,\varpi_0)_{_1}$ = (0°, 0.7, 0°) can be found when $e_p=0.3$. In the top of Fig. \ref{fig:2-1_3int_vs_3DHmaps} we present the results of three numerical integrations where the initial conditions were gradually being put further away from this ACR (always with $J=0$). The result is that the centre of libration starts to have a long-period oscillation that in the extreme case reaches almost 180° of amplitude. Besides, $e$ and $\varpi$ also have an important secular evolution. Therefore, this dynamical behaviour cannot be completely understood by means of the maps $\mathcal{R}(\sigma,\varpi)$, $\mathcal{R}(\sigma,e)$ or $\mathcal{R}(e, \varpi)$ because in all these maps, one of the three variables is considered fixed and besides, the topology of the contour curves changes whenever one of this magnitudes has a considerable variation. Hence, in this example we explore the entire $(\sigma,e,\varpi)$ space at once, plotting all the equilibrium points found with equation \ref{eq:dR_ds} and assigning different colours according the hamiltonian's value of each point. The colour assignment is vital to differentiate the $\mathcal{H}=C$ curves that will predict the secular evolution. This can be seen in the bottom of the Fig. \ref{fig:2-1_3int_vs_3DHmaps} where the comparisons where carried out between the numerical integrations (black curve) and the three-dimensional $\mathcal{H}$ surfaces. Some points of the surface´s edges were removed in order to have a better visualisation. This does not interfere with the analysis so far.

It is remarkable how good is the agreement between the numerical integrations and the model's surface for all the cases. Contrary to the $e_p=0.01$ case, here $\Delta e$ is much bigger when $e_i$ is displaced from $e_0$. Also $\varpi$ presents large secular variations when $\varpi_i$ is far away from $\varpi_0$.
With this method, we can observe which are the pathways that could increase greatly the asteroid´s eccentricity when is exactly in the 2:1 MMR with an eccentric perturber. An example of this is precisely the right one in the Fig. \ref{fig:2-1_3int_vs_3DHmaps} where the initial $e_i=0.32$ but after 200 kyrs it reaches values extremely close to 1. The surface bifurcation at low $e$ and $\sigma=\varpi=0$° is related to a separatrix we will analyse later in this section.

% FIGURA 9
\begin{figure*}
  \includegraphics[width=\textwidth]{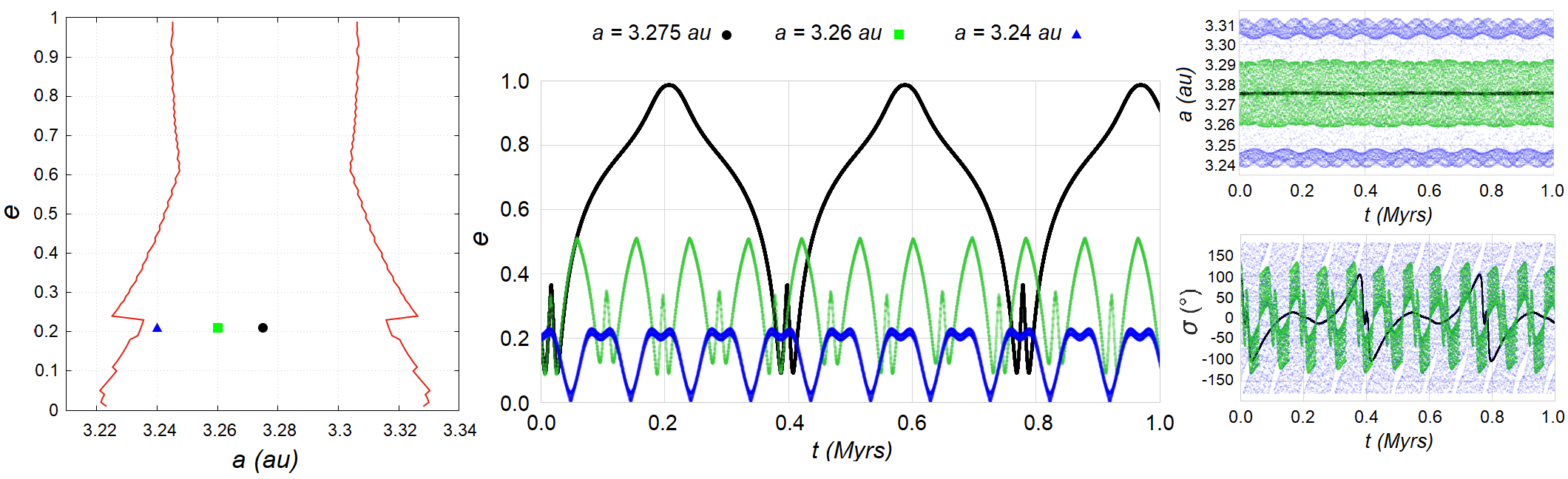}
  \caption{2:1 MMR with $e_p=0.3$. \textit{LEFT}: $(e, a)$ plane showing the width for $\varpi$=180°. The three coloured symbols shows the initial conditions for three numerical integrations. \textit{CENTRE}: $e(t)$ of the numerical integrations. \textit{RIGHT}: $a(t)$ and $\sigma(t)$ of the numerical integrations.}
  \label{fig:2-1_e2=0.3_int_borde}
\end{figure*}

Another surface exists (in the same panel) with an ACR peeking out in $(\sigma_0,e_0,\varpi_0)_{_2}$ = (180°, 0.99, 180°), but due to the high eccentricity value, all numerical integrations reached $e=1$ (as the contour curves do) in thousands of years, except if the initial conditions were exactly in the mentioned triplet. Consequently, is not a really important equilibrium point family from the practical point of view in this case. Nevertheless, we will see in the next example that this ACR point comes down in $e$ when $e_p$ increases and in fact, is the beginning of an entire ACR family.

There is another zone that is worth of being analysed. It is topologically in the same surface as the main ACR in $(\sigma_0,e_0,\varpi_0)_{_1}$ = (0°, 0.7, 0°) but is beyond of a separatrix, so the behaviour is quite different. In the Fig. \ref{fig:2-1_e2=0.3_separatriz} we show three different numerical integrations in green and the separatrix in black together with the $\mathcal{H}$ surfaces. In the first numerical integration the initial condition (yellow marker) is such that the contour curve still closes on itself and surrounds $(\sigma_0,e_0,\varpi_0)_{_1}$ (i.e. is still the first zone studied in fig. \ref{fig:2-1_3int_vs_3DHmaps}). The second initial condition is on the separatrix. The third one is beyond the separatrix, laying on a contour curve that suddenly vanishes for a higher $e$ value, therefore, is not a closed curve but an open one. In both last cases the amplitude of resonant libration is not zero anymore, being more pronounced in the third one. In this case, the numerical integration is more uncorrelated with the $\mathcal{H}$ surface, behaving more chaotically than the other two. 
Let's call this family of contour curves beyond the separatrix by \textit{open} curve family. In order to explain this behaviour, we have selected three points in the open curve family and plot $\mathcal{R}(\sigma)$ for each of them. The idea is to inspect what happens when $e$ increases in a single contour curve, approaching to the edge of this family (or surface´s edge). This can be seen in the Fig. \ref{fig:2-1_e2=0.3_3ptos} where the points are coloured differently. At the right of that figure are the $\mathcal{R}(\sigma)$ functions for each $(e,\varpi)$ pair, where an arrow with the same point´s colour indicates where is the resonant equilibrium point drawn in the $\mathcal{H}$ surface. 
Note how the resonant equilibrium point disappears ($\mathcal{R}_{\sigma\sigma}$ seems to decrease) as we move towards the edge. This happens when $e=0.75$ in this example. In that point, the system came from evolving adiabatically with $J=0$, but suddenly there is no more resonant equilibrium point, so, the "initial" condition in the green point is of $J>0$ because the particle now will librate around the nearest equilibrium point which is marked with a blue circle in that figure. This phenomenon of disappearing libration centres was already observed by \citet{2016CeMDA.126..369S} where some conveniently fragmented maps were presented to understand the secular evolution of a particle in the 1:11 MMR with Neptune.

% FIGURA 10
\begin{figure*}
  \includegraphics[width=\textwidth]{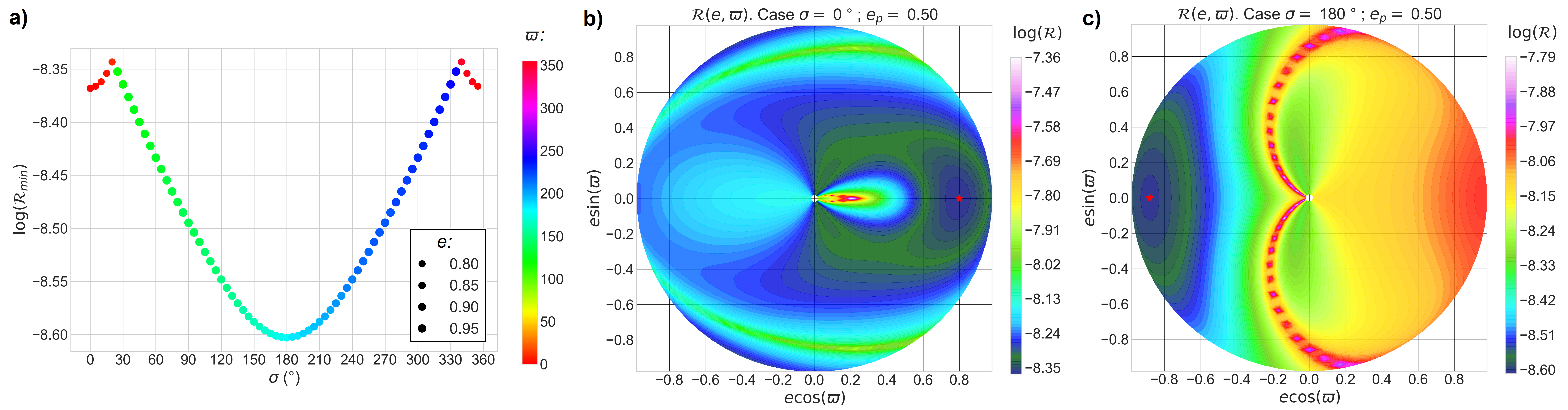}
  \caption{2:1 MMR with $e_p=0.5$. \textbf{a)} $\min{\{\mathcal{R}(e,\varpi)\}}$ vs $\sigma$. \textbf{b)} $R(e,\varpi)$ for $\sigma=0$° \textbf{c)} $R(e,\varpi)$ for $\sigma=180$°}
  \label{fig:2-1_e2=0.5_Rmin_sigma_Re1w1_s1=0_s1=180}
\end{figure*}

Finally we show in the Fig. \ref{fig:2-1_e2=0.3_int_borde} how the evolution is greatly modified if we move the particle from being in the exact MMR (in the closed curve family). We achieve this by just considering three numerical integrations with the same initial conditions except for $a$, which is displaced from the nominal value in two of them. The initial conditions for the particle are $e_i=0.21$, $\varpi_i$=180° and $\sigma_i$=102° with the following three $a_i$: 3.275 (nominal value), 3.26 and 3.24 au. In the mentioned figure it can be observed the resonant structure in the $(a,e)$ plane, where the resonant width was calculated using the formula derived in \citet{2020CeMDA.132....9G}. In the same plane, there are three symbols indicating the initial values for the numerical integrations. We also show the temporal evolution of $e$, $a$ and $\sigma$. In the first case the particle is in exact MMR with a similar evolution of the third case shown in Fig. \ref{fig:2-1_3int_vs_3DHmaps}, where the eccentricity is excited almost to 1. The centre of libration evolves in the secular time-scale but the resonant libration amplitude is zero. In the second case, there is a different evolution for $e$ with higher frequency and a lower secular amplitude. As expected, $a$ is centred in the nominal value but with a non-zero resonant amplitude of libration. $\sigma$ has also a non-zero amplitude of libration with higher secular frequency (as $e$). In the third case $a$ enters in a stickiness behaviour proper of being at the edge of the resonance, $\sigma$ alternates between circulating and librating whereas $e$ has a completely different evolution with much lower secular amplitude. 

In this example (2:1 MMR with $e_p=0.3$) we notice how different could be the secular evolution between being in deeply resonant motion and in the edge of the resonance or in non-resonant motion. In particular, this mechanism (of being in deep MMR) could be responsible for generating extremely high eccentricity orbits. As a final observation we should emphasise the importance of the $J=0$ hypothesis in order to the model predict reliably the dynamical evolution.

\subsubsection{Very high-eccentricity planet ($e_p=0.5$)}
\label{sec_ep=0.5}

The third example we present for this resonance is with $e_p=0.5$. In the Fig. \ref{fig:2-1_e2=0.5_Rmin_sigma_Re1w1_s1=0_s1=180}a) can be seen the $\min{\{\mathcal{R}(e,\varpi)\}}$($\sigma$) where this time there are two minimums, one in $\sigma = 0$° and the other in $\sigma = 180$°. The first one occurs for $(e, \varpi) = (0.8, 0^{\circ})$ whereas the second one for $(e, \varpi) = (0.88, 180^{\circ})$, as can be observed in the same Fig. \ref{fig:2-1_e2=0.5_Rmin_sigma_Re1w1_s1=0_s1=180}b) and c). The red star in these $\mathcal{R}(e,\varpi)$ maps marks the minimum. 
If we compare numerical integrations with these $\mathcal{R}(e,\varpi)$ maps when the initial conditions are in the $\varpi$ libration zone, we have good agreement. In these zones $e$ presents moderate variations whereas $\varpi$ presents low variations.

In the $\sigma=0$° case, if the initial condition is $(e_i,\varpi_i)=$ (0.61,0°), it means that the particle is on the separatrix. In that situation, the centre of resonant libration starts to circulate after approximately 30 kyrs and the map loses validity. To explain this behaviour, we should make use of the $\mathcal{H}$ surfaces described in the section \ref{sec:H_maps}. Nevertheless, due to the complexity of this example, several snapshots of the $\mathcal{H}$ surface were taken from different view angles which can be observed in the appendix (Fig. \ref{fig:2-1_e2=0.5_int_complicado}) together with the numerical integration. This is another example of high eccentricity variations. Note how a particle with $e_i \sim 0.2$ could reach values close to 1, if $\varpi_i$ and $\sigma_i$ are properly selected. Observe how despite the intricate behaviour of the evolution, the numerical integration follows this three-dimensional contour curves in the $\mathcal{H}$ surface. In practice is convenient to use an interactive 3D graphic manipulator (for example ipyvolume from pyhton library) to inspect the dynamical structure more easily. 

% FIGURA 11
\begin{figure*}
  \includegraphics[width=0.99\textwidth]{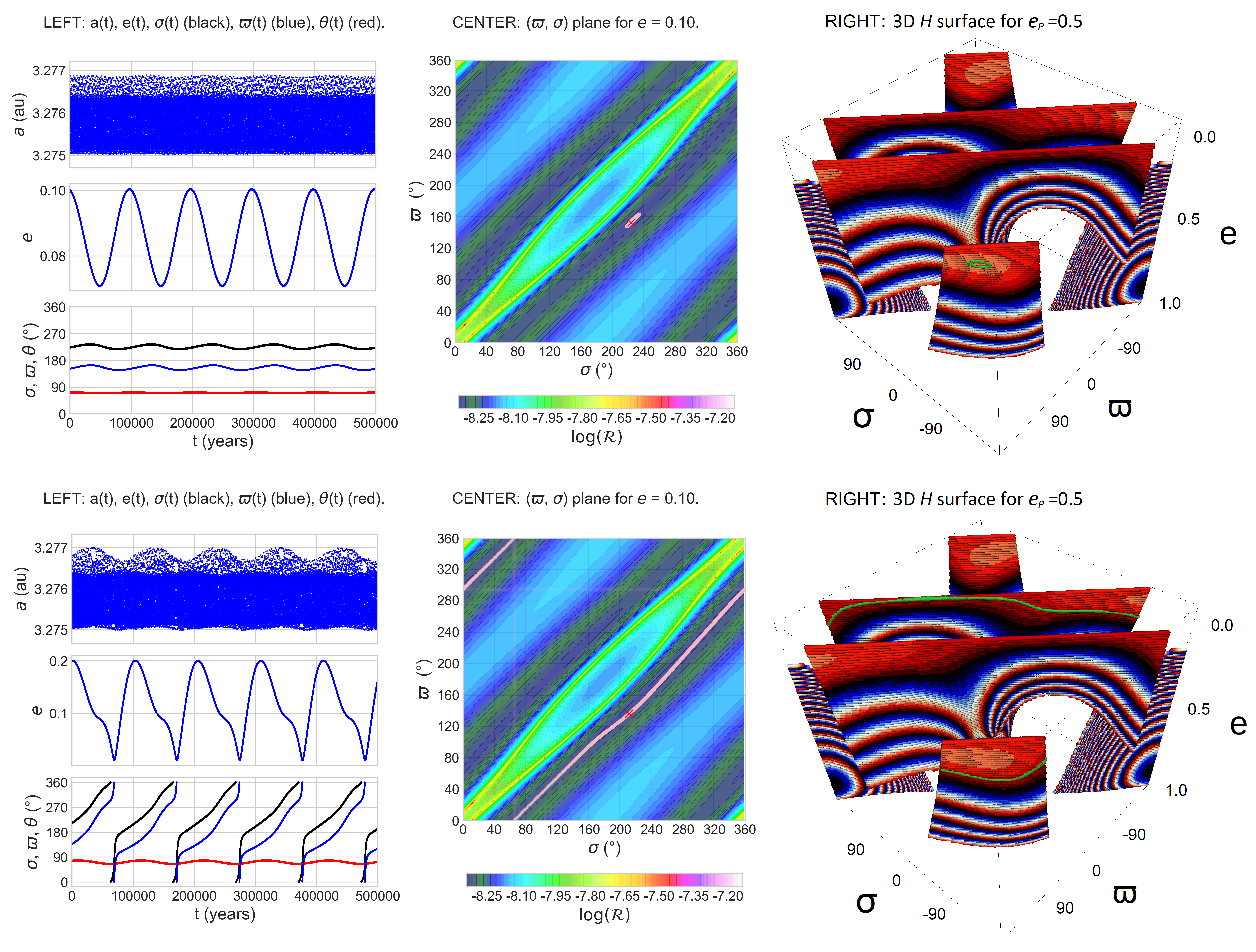}
  \caption{2:1 MMR with $e_p=0.5$. Two numerical integrations compared with $\mathcal{R}(\varpi,\sigma)$ maps and $\mathcal{H}$ surfaces. Two asymmetric ACRs are found at $(\sigma_0, e_0, \varpi_0)$=($\mp$133°, 0.08, $\pm$155°). \textit{TOP}: $(\sigma_i, e_i, \varpi_i)$=(-144°, 0.10, 142°). \textit{BOTTOM}: $(\sigma_i, e_i, \varpi_i)$=(-148°, 0.20, 136°).}
  \label{fig:2-1_int_e2=0.5_RW1s1_3DH}
\end{figure*}

Finally, in the top of the Fig. \ref{fig:2-1_int_e2=0.5_RW1s1_3DH} is shown a case where $\sigma$ is librating in the asymmetric angle of 227°, which is very interesting because for this interior resonance, it was thought that these asymmetric librations did not exist or exists but for $m_p/m \sim 1$ \citep{2003ApJ...593.1124B}. %,2005dpps.conf....3B} 
In \citet{BMF2006} they extended the search of corotational solutions for higher eccentricities values but did not appear asymmetric points for $m_p>m$ in the masses range they explored. Despite this, in the fig. 2 of their work there is a zone near $e_p=0.5$ and $e=0.1$ where this point could asymptotically exist. 
In the Fig. \ref{fig:2-1_int_e2=0.5_RW1s1_3DH} is also shown the $\mathcal{R}(\sigma,\varpi)$ map and the 3D $\mathcal{H}$ surface, both with the numerical integration overlapped. Both maps predict this asymmetric ACR as can be seen. Moreover, they predict another asymmetric ACR point located symmetrically opposite with respect to the origin in the $(\sigma,\varpi)$ plane. 
Despite this interesting feature discovered in the dynamics, it is noteworthy that this ACR seems to be pretty weak because of its limited extension in the space (contrary to the main ACR point found in the $e_p=0.3$ case) which can be proved if we see the bottom part of the Fig. \ref{fig:2-1_int_e2=0.5_RW1s1_3DH}. Here the initial conditions are slightly changed, resulting in a completely different behaviour of both, $\sigma$ and $\varpi$. They circulate, as the contour curves on the $\mathcal{H}$ surface predicts, whereas $\theta$ is approximately fixed. Note how in both numerical integrations $e$ does not change considerably, so, with the $\mathcal{R}(\sigma,\varpi)$ map would have been enough for explaining the encountered behaviour.

If we compare the $\mathcal{H}$ surface's topology between this case and the $e_p=0.3$ case, we can point out for example that in the $e_p=0.5$ case, the bifurcation occurs for a higher $e$ value, making it bigger. Secondly, the surface has a more complicated topology than in the $e_p=0.3$ case. This implies that in the former case, $\sigma$ secular variations will be bigger when the initial conditions are displaced from the ACRs. Finally, a more eccentric perturber produced a second main equilibrium point in $\sigma=\varpi=180$° with more dynamical relevance (recall the "instability" surrounding this point when $e_p=0.3)$. In the Table \ref{tab:2-1_sec_eq_points} are summarized all the ACRs found in the 2:1 MMR for the particular $e_p$ values studied. 

\begin{table}
	\centering
    \resizebox{0.48\textwidth}{!}{\begin{tabular}{|l|llllll|}
    \hline
     $(\sigma_0,e_0,\varpi_0)$  & 1   & 2    & 3, 4        \\ \hline%\cline{2-7} 
        $e_p=0.01$            & (0°, 0.73, 0°) &   -  & -  \\
        $e_p=0.3$             & (0°, 0.7, 0°) & (180°, 0.99, 180°) & -  \\ 
        $e_p=0.5$             & (0°, 0.79, 0°) & (180°, 0.88, 180°) & ($\mp$133°, 0.08, $\pm$155°)  \\ \hline
    \end{tabular}}
    \caption{Summary of ACRs in the 2:1 MMR for the three cases investigated.}
    \label{tab:2-1_sec_eq_points}
\end{table}

\subsubsection{ACR families}
\label{sec:ACRs}

\begin{figure*}
  \includegraphics[width=0.95\textwidth]{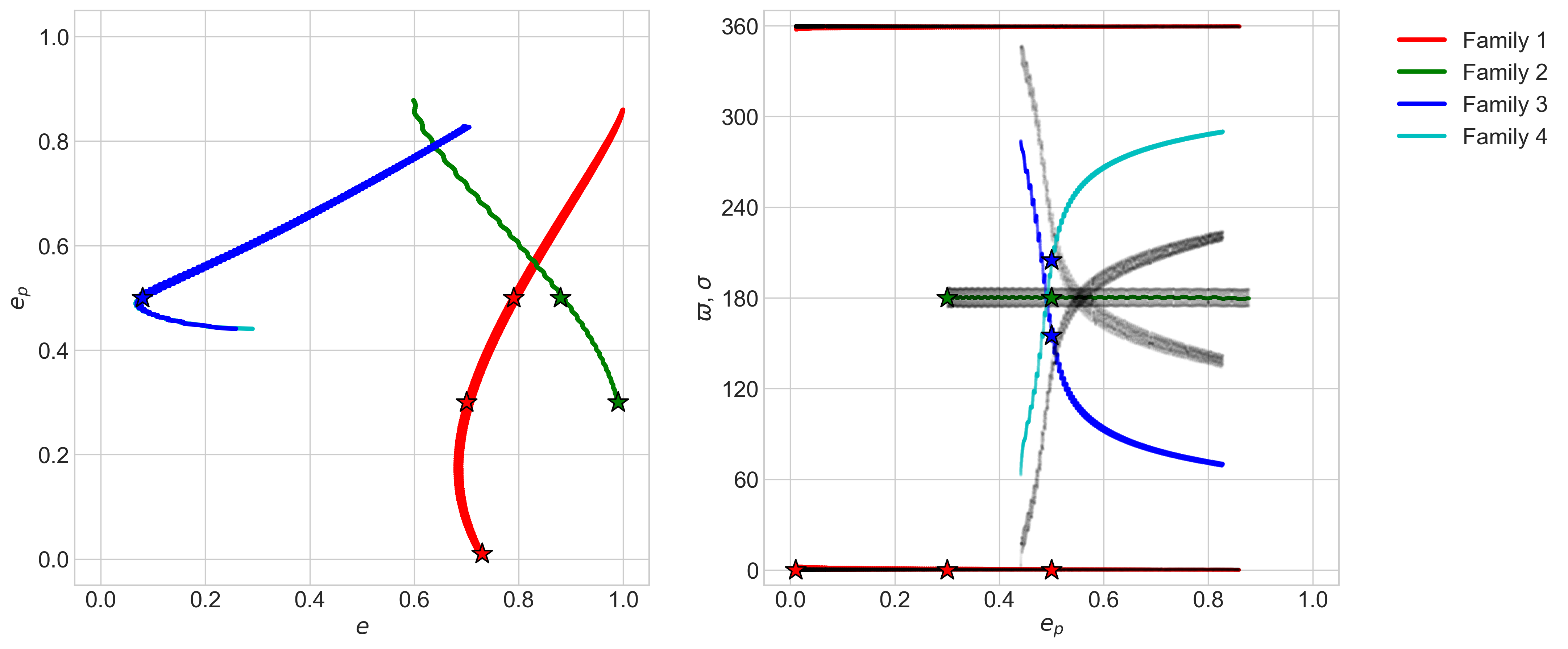}
  \caption{ACR families for the 2:1 MMR. Families 1 and 2 are the symmetric ones whereas families 3 and 4 are the asymmetric ones. In the right handed panel $\varpi$ has the same color code as $e$ and $\sigma$ is in black. The stars indicate the ACRs found in the cases studied in detail in previous sections (see Table \ref{tab:2-1_sec_eq_points}).}
  \label{fig:2-1_ACRs}
\end{figure*}

In order to be aware of the entire families where the ACR points of the Table \ref{tab:2-1_sec_eq_points} belongs, we explore the space for any $e_p$ value up to ~0.85. The result can be seen in Fig. \ref{fig:2-1_ACRs}. For $e_p<0.3$ only exist the $\sigma=\varpi=0$° symmetric ACR family. From that point, appears another symmetric family at $\sigma=\varpi=180$°. At $e_p\simeq0.44$ starts the third and fourth families, which are asymmetric. In this family, not only $e$ changes for different $e_p$ values, but also $\varpi$ and $\sigma$.

% FIGURA 12

\subsection{Application: The Planet 9}
\label{sec:planeta9}

\begin{figure}
  \includegraphics[width=\linewidth]{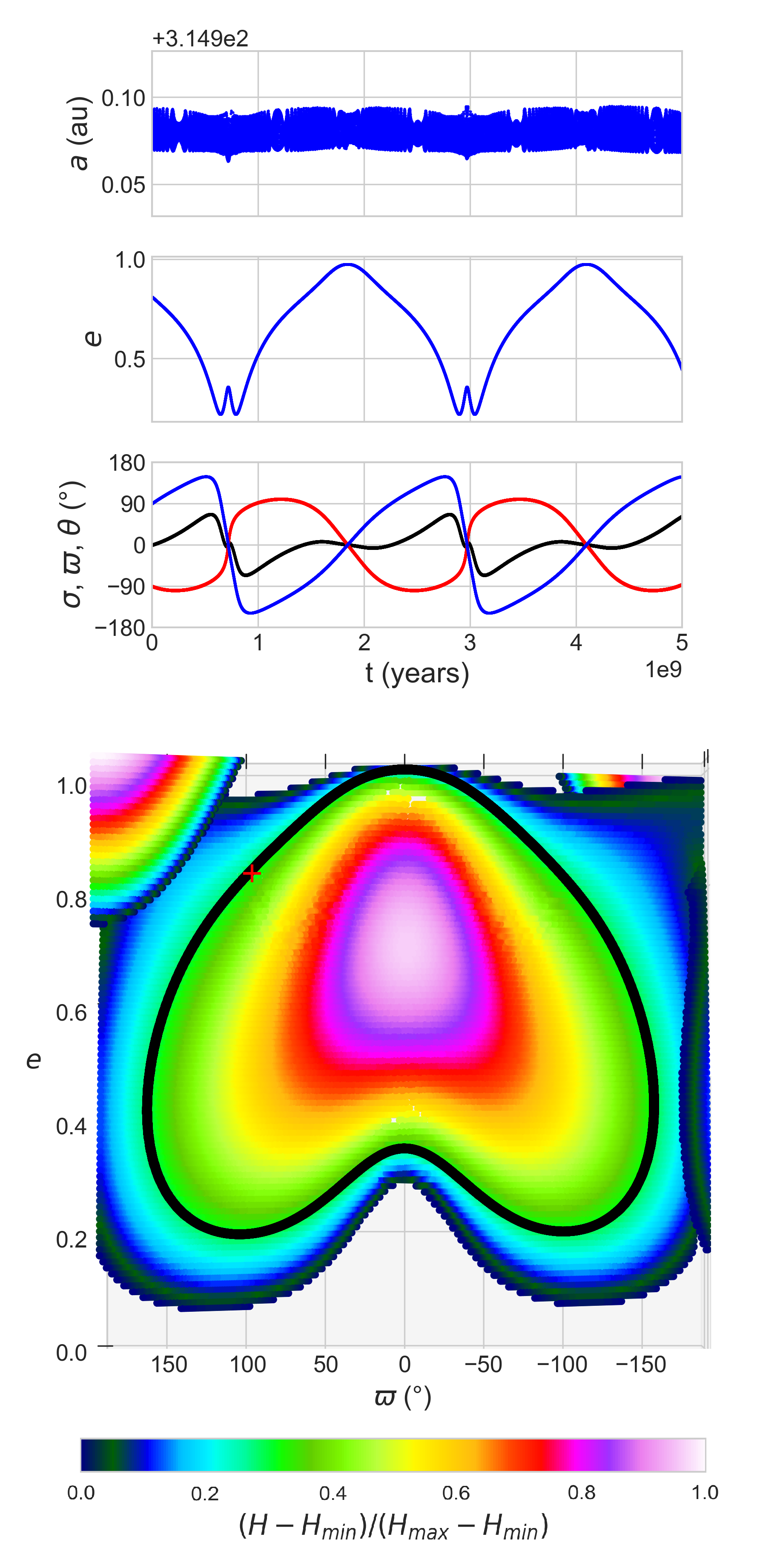}
  \caption{2:1 MMR with $e_p = 0.25$, which is the Planet 9 canonical eccentricity. \textit{TOP:} Numerical integration which initial conditions are: $(\sigma_i, e_i, \varpi_i)$=(0°, 0.81, 90°). \textit{BOTTOM:} $\mathcal{H}$ surface projection onto the $(e, \varpi)$ plane with the numerical integration in black. The red cross indicates the initial conditions. }
  \label{fig:2-1_int_planet9_3DH}
\end{figure}

In this section we apply the method to the hypothetical Planet 9, using for this its canonical orbital elements values and a mass of $m_p=5$ $m_{\earth}$ \citep{BATYGIN20191}, but taking, without losing generality, $\varpi_p=0$° and $i_p=0$°, since we are not considering the Solar System planet's.
The objective here is to confirm that low-to-high eccentricity pathways can exists and could be a partial explanation of those high-eccentric distant TNO observed nowadays, assuming that they are in MMR with the Planet 9. 
First of all, we select the 2:1 MMR and construct the 3D $\mathcal{H}$ surface for the Planet 9. Then we do a projection of it in the $(e, \varpi)$ plane which can be seen in the Fig. \ref{fig:2-1_int_planet9_3DH} (observe the similarity with the map of Fig. \ref{fig:2-1_3int_vs_3DHmaps}, due to the $e_p$ values are pretty similar). To do this, we had to filter some of the equilibrium points to have a more complete visualisation of the region of interest. Observe how this projection could be very tricky if not impossible when the $\mathcal{H}$ surface become more complicated, as in the $e_p=0.5$ case. There, the projection would deprive of the necessary information to understand the secular behaviour of $\sigma$.

In the same Fig. \ref{fig:2-1_int_planet9_3DH} we compare with a numerical integration of 5 Gyrs, where a particle suffer extreme changes in eccentricity. This confirms that low-to-high eccentricities pathways exists with really stable orbits, assuming no inclination respect Planet 9's orbit, which for TNOs could be a bit restrictive hypothesis given the different inclinations and longitudes of nodes these objects have. This application study does not pretend to explain exhaustively the high eccentric orbits present in these distant ($a$>250 au) TNO population but just illustrate that at least a secular evolution inside this MMR could be partially responsible for some of the orbital characteristics. In our example, the particle has $a=314.98$ au (in order to be in the exact MMR), a value close to the semi-major axis of the high-eccentricity objects 2015 GT50, 2004 VN112 and 2014 SR349.

Historically was proposed that there should exist an anti-alignment between distant TNO and Planet 9 peri-centres \citep{BB2016}. However, we found that a peri-centre alignment could be another option, as the ACR is in $\varpi=0$. Furthermore, the $\mathcal{H}=C$ curves shrinks toward $\varpi=0$ when $e\rightarrow 1$, which could result in the following scenario: a set of fictitious initial low-eccentric particles with non-aligned pericenter (the lack of equilibrium points there is due to some filtering that was done in order to avoid bifurcation zone, see Fig. \ref{fig:2-1_3int_vs_3DHmaps}) could increase $e$ and converge to approximately the same $\varpi$, producing the known perihelion clustering \citep{Trujillo2014}.

As we mentioned, this example does not pretend to be an exhaustive and deep study of the Planet 9 hypothesis. It is just to illustrate that, in case of more distant TNO being discovered, it would be expected a peri-apsis alignment in those objects locked in the 2:1 MMR. Naturally, other MMR should be considered to make a more complete analysis. Nevertheless, no enough distant TNO have been observed to obtain a clear signature, in the semi-major distribution, that suggest that the resonant mechanisms are dominant in this population. Some extra analysis is required to understand up to what extent is important the secular mechanisms inside MMR in relation with these mechanisms outside MMR. This is not fully understood as both mechanisms are capable of producing similar orbital excitations. In \citet{2016A&A...590L...2B} there is an interesting discussion about this issue where the different (dis)advantages of both mechanisms are commented.
 
 % FIGURA 13

\section{Discussion and conclusions}

We have developed a simple but useful technique, based on \citet{2020CeMDA.132....9G}, that allow us to obtain the secular evolution of any zero-amplitude resonant asteroid being perturbed by a coplanar massive body for arbitrary values of $e$ and $e_p$. In particular predicts the secular evolution of $e$, $\varpi$ and $\sigma$, i.e., the resonant libration centre. 
Some of the advantages of this method are listed below:
\begin{itemize}
    \item Allows to find every ACR in the entire $(\sigma,e,\varpi)$ space.
    \item Predicts correctly the secular evolution of $e$, $\varpi$ and $\sigma$ in the zero-amplitude libration regime, for any initial condition, including those far away from ACRs.
    \item Allows to seek for dynamical paths which could increase greatly $e$ (see Fig. \ref{fig:2-1_3int_vs_3DHmaps} for example).
    \item Allows to find the separatrixes that trigger a behavioural change in $\varpi$ between librating and circulating when the centre of libration is fixed (see c) and d) examples in Fig. \ref{fig:2-1_e2=0.01_int_Re1w1_Ha1s1}).
    \item Allows to find separatrixes in the $(\sigma,e,\varpi)$ space, that could delimit open contour curves (unstable) from closed contour curves families (stable).
    \item There are no limitations for $k$, $k_p$, $e$ and $e_p$.
\end{itemize}

In general the complexity of the dynamical behaviour increased for larger $e_p$ values, requiring the utilisation of the 3D $\mathcal{H}$ surfaces. Basically, on one hand, if the centre of libration has negligible variations, with the contour curves in the $(e\sin{\varpi},e\cos{\varpi})$ plane is possible to analyse the secular evolution. On the other hand, when $e_p$ is large enough, $\sigma$ could start to vary considerably in the secular time-scale, requiring a more sophisticated way of representing the phase space.

With respect to the examples examined here, there are some important remarks to be done. First of all, for the 2:1 MMR quasi-circular case, it was found an ACR point in ($\sigma_0$, $e_0$, $\varpi_0$) = (0°, 0.73, 0°) which is a similar result as the obtained in \citet{2017A&A...605A..23P} (see fig. 3a in their work). In this case, the secular evolution of $e$ and $\varpi$ was correctly predicted by our model as long as the $\varpi$ librating/circularising regime limit. In particular, we note that the maximum $e$ variation in the $\varpi$ libration regime is $\Delta e=0.22$ whereas in the $\varpi$ circulation regime, $e$ variations are at most of $\Delta e\simeq0.1$, in agreement with the model.

In the high eccentricity case ($e_p=0.3$) things get more interesting. The centre of resonant libration starts to evolve in secular time-scale, following the contour curves of the $\mathcal{H}$ surface. As in the previous case there is one main ACR, almost in the same place but with the slightly different eccentricity of $e=0.7$. There is also another ACR at $e=0.99$ and due to this extreme value, any minimal displacement from that point results in a short evolution because the particle reaches $e=1$ (as the contour curves predict) rapidly. Both ACR points also seems to be present in the results of \citet{2017A&A...605A..23P} (see fig. 6b in their work). 

In the very high eccentricity case ($e_p=0.5$) of this MMR, there are two main ACR points. The first one at ($\sigma$, $e$, $\varpi$) = (0°, 0.8, 0°) whereas the second one at ($\sigma$, $e$, $\varpi$) = (180°, 0.88, 180°). There are also two weaker asymmetric ACR points which are at ($\sigma$, $e$, $\varpi$) = (227°, 0.08, 155°) and ($\sigma$, $e$, $\varpi$) = (155°, 0.08, 227°). Numerical integrations are contrasted with this asymmetric point and we conclude that they exists but with a really narrow secular libration width because separatrixes are too close. Therefore, they are less relevant than the others.

ACRs complete families were determined with a full exploration varying continuously $e_p$ up to 0.85. The higher $e_p$ is, the more families coexists. In addition to the location of all the ACRs, some pathways that increase $e$ greatly have been found and tested, in the specific cases of moderate and high eccentric perturber ($e_p=0.3$ and $e_p=0.5$).

We made a detailed analysis to understand the bifurcation at low $e$ (present almost in all the MMR for $e_p \gtrsim 0$) in the $\mathcal{H}$ surface. This bifurcation coincides with a separatrix that divides two different contour curve families, one with closed contour curves that surrounds the main ACR and other with open contour curves. The last one is related to the $J=0$ hypothesis breakdown, due to the discontinuity in the contour curves formed by the resonant equilibrium points. 

We also numerically compared the evolutions between being in the exact MMR and displaced from it. The results (Fig. \ref{fig:2-1_e2=0.3_int_borde}) allow to conclude that the secular evolution could be very different when being in a deep MMR than when not. In particular, in deep MMR can exist dynamical paths that could lead to extremely large changes in some orbital elements, for example in $e$.

Finally, an application was presented regarding the hypothetical Planet 9 and a mass-less object in 2:1 MMR with it. This could help with the explanation of those high-eccentric distant TNO observed in the last decades. If any of them was effectively excited through this mechanism, they should have remained in MMR with the Planet 9 during the Gyr time-scale. Provided of this, its $\varpi$ should be pretty similar to the Planet 9's one because of the particular shapes of the $\mathcal{H}=C$ curves on the $\mathcal{H}$ surface.

In the future some applications of this method could be used for understanding the secular evolution of resonant exocomets and exoasteroids perturbed by eccentric exoplanets. It could be useful also in the understanding of high eccentric exoplanetary systems where one of the planets has negligible mass compared to the other.

An extension of this method to the spatial problem could be done without any extra theoretical limitations. The unique drawback is the impossibility of having all the dynamical features in one single plot, as we have in the planar case with the $\mathcal{H}$ surfaces. A possible approach to overcome this issue could be generate several $\mathcal{H}$ surfaces for different $(i, \Omega)$ pairs and obtain results from there. Another option could be to implement an algorithm to extract relevant information (without doing a single plot) as for instance the location of the ACRs, the number of equilibrium points families, separatrixes information, other libration islands, etc.

Some extra examples in the 3:1 and 3:2 MMRs where documented in the appendixes just to illustrate interesting secular evolutions which can also be explained with this technique.

\section*{Acknowledgements}

We want to acknowledge the ANII and PEDECIBA support that were fundamental to make this study possible. We are also grateful with Cristian Beaugé because of his valuables contributions through discussions and facilitation of some software tools. Finally, we appreciate the corrections given by Dr. Hanlun Lei which were very useful to improve the final manuscript.

% The Acknowledgements section is not numbered. Here you can thank helpful
% colleagues, acknowledge funding agencies, telescopes and facilities used etc.
% Try to keep it short.

%%%%%%%%%%%%%%%%%%%%%%%%%%%%%%%%%%%%%%%%%%%%%%%%%%
\section*{Data Availability}

The data underlying this article is available in astronomy department website of Facultad de Ciencias, UdelaR at http://www.astronomia.edu.uy/repositoryPonsGallardo/, and can be accessed through the link.

\newpage
%%%%%%%%%%%%%%%%%%%% REFERENCES %%%%%%%%%%%%%%%%%%

\bibliographystyle{mnras}
\bibliography{referencias} % if your bibtex file is called example.bib

%%%%%%%%%%%%%%%%%%%%%%%%%%%%%%%%%%%%%%%%%%%%%%%%%%
%\newpage
%%%%%%%%%%%%%%%%% APPENDICES %%%%%%%%%%%%%%%%%%%%%

\appendix

\section{2:1 MMR - \lowercase{$e_p=0.5$} complex case}

\begin{figure*}
  \includegraphics[width=\textwidth]{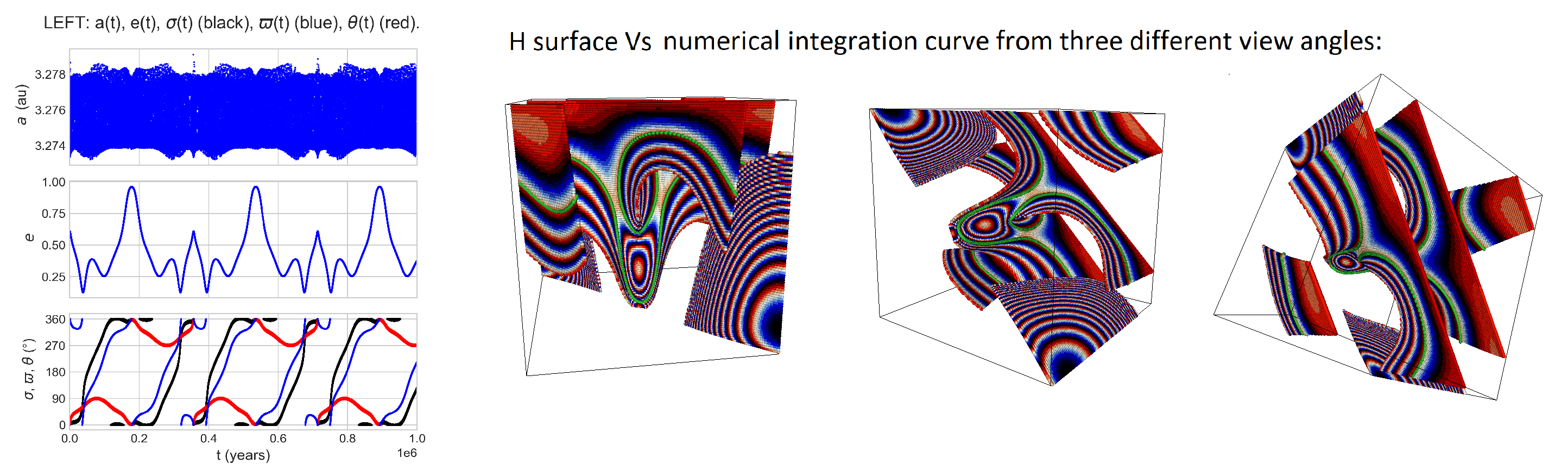}
  \caption{2:1 MMR with $e_p = 0.5$. \textit{LEFT}: $a_1(t)$, $e(t)$, $\sigma(t)$ (black), $\varpi(t)$ (blue) and $\theta(t)$ (red) from the numerical integration with $(\sigma_i,e_i,\varpi_i)$=(0°, 0.61, 0°). \textit{RIGHT}: $\mathcal{H}$ surfaces from different angles. In green is the numerical integration.}
  \label{fig:2-1_e2=0.5_int_complicado}
\end{figure*}

In this appendix section, we show a complex evolution case when $e_p=0.5$. The comparison with the numerical integration is shown in \ref{fig:2-1_e2=0.5_int_complicado} where some snapshots of the $\mathcal{H}$ surface can be seen. Note the really intricated secular evolution which occurs when the initial condition is $(\sigma_i,e_i,\varpi_i)$=(0°, 0.61, 0°). This point turns out to be just out the $\varpi$ libration zone, i.e., is beyond the separatrix over the horizontal axis, as can be seen in the map $\mathcal{R}(e,\varpi)$ for $\sigma=0$° in the Fig. \ref{fig:2-1_e2=0.5_Rmin_sigma_Re1w1_s1=0_s1=180}.

% FIGURA A1

\section{3:1 MMR example}
\label{sec:3-1}

\begin{figure}
  \includegraphics[width=0.92\linewidth]{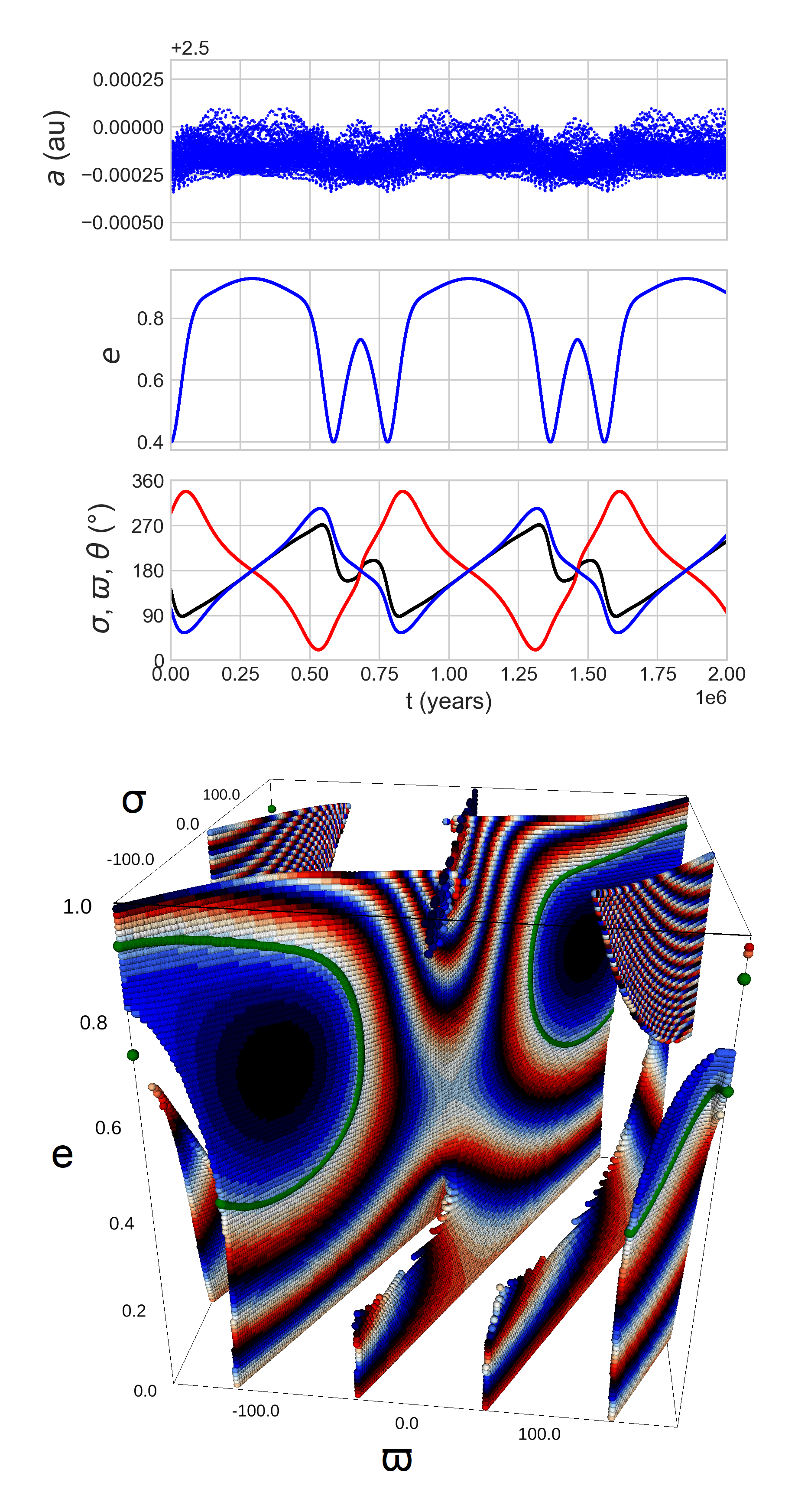}
  \caption{3:1 MMR with $e_p = 0.3$. \textit{TOP:} Numerical integration which initial conditions are: $(\sigma_i, e_i, \varpi_i)$=(143°, 0.40, 104°). \textit{BOTTOM:} $\mathcal{H}$ surface with the integration in green.}
  \label{fig:3-1_int_3DH}
\end{figure}

In this appendix section, we present an example just to illustrate an interesting behaviour found around two asymmetric ACR points. This behaviour can be observed in the Fig. \ref{fig:3-1_int_3DH}. In this example the perturber has $e_p=0.3$ and the ACR are located at $(\sigma,e,\varpi)$=(135°, 0.65, 101°) and $(\sigma,e,\varpi)$=(225°, 0.65, 259°), being both points symmetrical to each other with respect to the origin in the $(\sigma,\varpi)$ plane, as happened in the 2:1 MMR with $e_p=0.5$. In this case the initial conditions are such that the secular behaviour corresponds to an alternation between circumnavigating one ACR and the other. This result in a very complicated temporal evolution for $e$, $\varpi$ and specially for $\sigma$. The adequate way to fully explain this is with the $\mathcal{H}$ surface, because just with the other maps or inspecting $\sigma(t)$, $e(t)$ and $\varpi(t)$, there is no enough information to understand the behaviour. Note how for low $e$, there is a bifurcation and also some open family curves can be observed.

% FIGURA B1

\section{3:2 MMR example}
\label{sec:3-2}

\begin{figure}
  \includegraphics[width=0.92\linewidth]{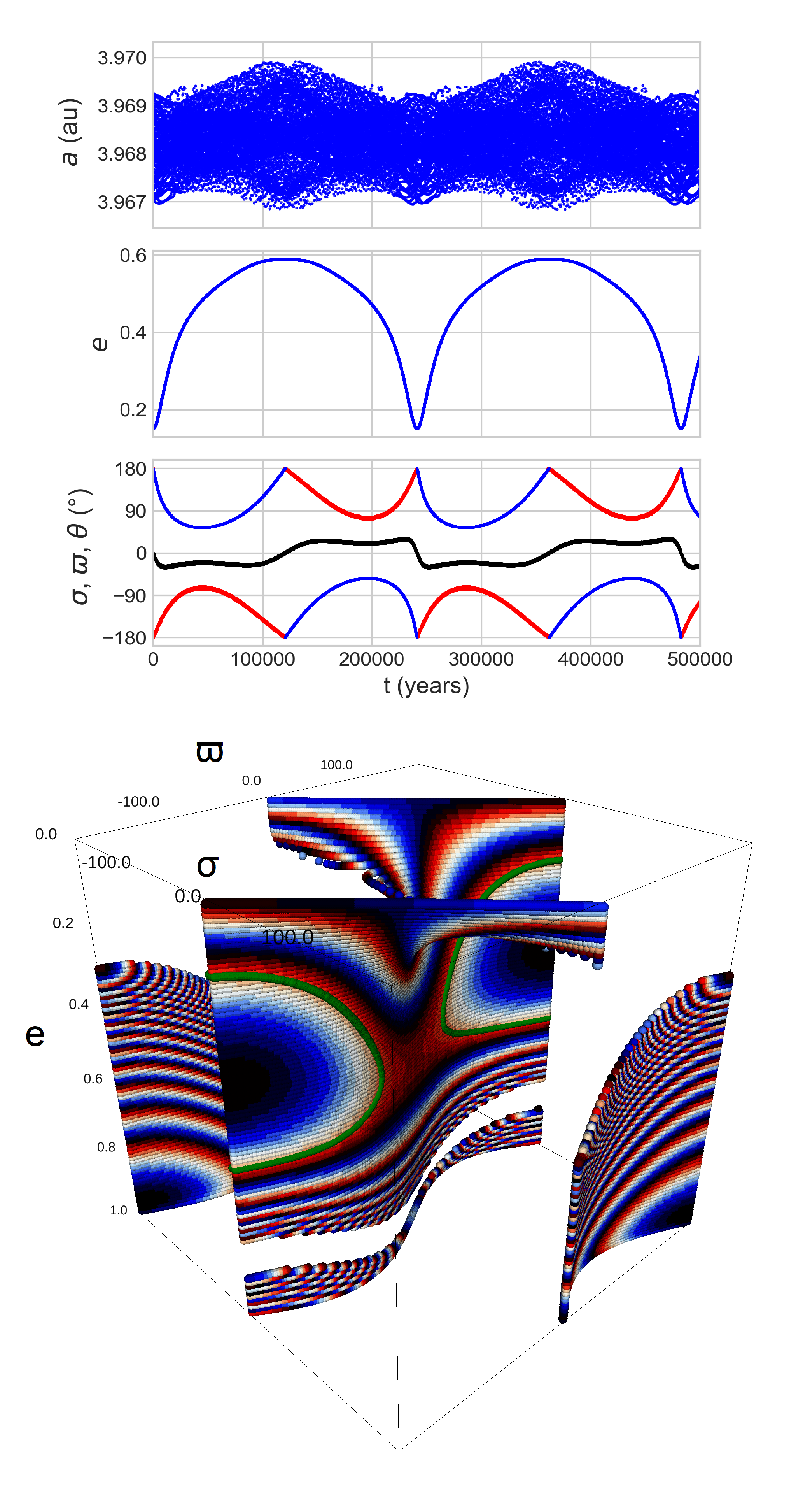}
  \caption{3:2 MMR with $e_p = 0.1$. \textit{TOP:} Numerical integration which initial conditions are: $(\sigma_i, e_i, \varpi_i)$=(0°, 0.15, 180°). \textit{BOTTOM:} $\mathcal{H}$ surface with the numerical integration in green.}
  \label{fig:3-2_int_3DH}
\end{figure}

In this appendix section, we also present an example just to illustrate an interesting behaviour found in $\sigma$. This behaviour can be observed in the Fig. \ref{fig:3-2_int_3DH}. In this example the perturber has $e_p=0.1$ and in this situation exists a main ACR point in $(\sigma,e,\varpi)$=(0°, 0.38, 180°). In this case the initial conditions are such that the secular behaviour of $\sigma$ results in a rectangular wave in time. This could be interpreted as a bi-stable situation because for the most of the time, $\sigma$ seems to librate in $\sim$ 25° and then changes rather fast to librate in $\sim$ -25°, similar to the behaviours found by \citet{2006Icar..181..205G}. However, there are not asymmetric libration points here but the behaviour is due to the $\mathcal{H}$ surface´s shape in the $(\sigma,e,\varpi)$ space and the particular initial conditions chosen. Once more we can see the bifurcation for low $e$ and other two ACR at $\sigma=\varpi=180$ °, one for $e\simeq0.35$ and the other in $e\simeq0.95$, both being pretty close to the edge of this second surface. 
 
% FIGURA C1

%%%%%%%%%%%%%%%%%%%%%%%%%%%%%%%%%%%%%%%%%%%%%%%%%%

% Don't change these lines
\bsp	% typesetting comment
\label{lastpage}
\end{document}